\documentclass{article}

\usepackage{PRIMEarxiv}

\usepackage[utf8]{inputenc} % allow utf-8 input

\UseRawInputEncoding

\usepackage[T1]{fontenc}    % use 8-bit T1 fonts
\usepackage{hyperref}       % hyperlinks
\usepackage{url}            % simple URL typesetting
\usepackage{booktabs}       % professional-quality tables
\usepackage{amsfonts}       % blackboard math symbols
\usepackage{enumitem} % remove indent fromo list
\usepackage{nicefrac}       % compact symbols for 1/2, etc.
\usepackage{microtype}      % microtypography
\usepackage{lipsum}
\usepackage{graphicx}
\usepackage{amssymb}
\graphicspath{{Figures/}}     % organize your images and other figures under media/ folder
\usepackage{fancyvrb}
\usepackage{pdfpages}

\usepackage[normalem]{ulem}

\usepackage{setspace}
\usepackage[switch]{lineno}
\title{Uncovering the Size of the Illegal Corporate Service Provider Industry in the Netherlands: a Network Approach}

\author{
  Javier Garcia-Bernardo \\
  Department of Methodology \& Statistics \\
  Utrecht University, the Netherlands\\
  \texttt{javier.garcia.bernardo@gmail.com}
  \And
  Joost Witteman, Marilou Vlaanderen \\
  SEO Amsterdam Economics \\
  Amsterdam, the Netherlands\\
  \texttt{\{j.witteman, m.vlaanderen\}@seo.nl}
}

\begin{document}
\maketitle

\begin{abstract}
Economic crimes such as money laundering, terrorism financing, tax evasion or corruption almost invariably involve the use of a corporate entity. Such entities are regularly incorporated and managed by corporate services providers (CSPs). Given this potential for enabling economic crime, the CSP industry in the Netherlands is heavily regulated and CSPs require a license to operate. Operating without a licence is illegal. 
In this paper, we estimate the size of the illegal CSP sector in the Netherlands.
For this, we develop a classification method to detect potentially illegal CSPs based on their similarity with licensed CSPs. Similarity is computed based on their position within the network of directors, companies and addresses, and the characteristics of such entities.
We manually annotate a sample of the potential illegal CSPs and estimate that illegal CSPs constitute 31--51\% of the total number of CSPs and manage 19--27\% of all companies managed by CSPs.
Our analysis provides a tool to regulators to improve detection and prevention of economic crime, and can be extended to the estimation of other illegal activities. 

\end{abstract}

% keywords can be removed
\keywords{economic crime \and network analysis \and corporate service providers \and Netherlands }

\section{Introduction}
%Importance
Economic crimes such as money laundering, terrorism financing, tax evasion or corruption almost invariably involve the use of a corporate entity \cite{oecdCorporateVeilUsing2001}. Such entities can be directly incorporated and managed by the ultimate beneficial owner, or alternatively, the incorporation and management of corporate entities can be delegated to corporate service providers (CSPs, `Trustkantoren' in Dutch) \cite{oecdCorporateVeilUsing2001, fatf-egmontgroupConcealmentBeneficialOwnership2018}.

The Netherlands has a significant CSP industry \cite{commissiedoorstroomvennootschappenOpWegNaar2021}.
The use of CSPs is usually legitimate, but without proper control, CSPs may provide opportunities (often unintentionally) to their clients to conceal their identity and/or the nature of their activities \cite{oecdCorporateVeilUsing2001,fatf-egmontgroupConcealmentBeneficialOwnership2018}. 
Given the potential of CSPs to enable economic crime  \cite{de2017role,watervalHoeWitwasserGepakt2021, degrootAfrikaRijksteVrouw2021, degrootTrustcowboysZonderVergunning2021}, the
provision of CSP services requires a license and is subject to supervision by the Dutch central bank (``De Nederlandsche Bank'', DNB)\footnote{One of the responsibilities of licensed CSPs is to monitor their clients and detect potential economic crimes. If licensed CSPs fail to adequately monitor their clients, they risk losing their license \cite{dnbDNBHeeft20192021, dnbDNBGeeftAanwijzing2021}.}.  In the Netherlands, the Act on the Supervision of Trust Offices 2018 (``Wet toezicht trustkantoren''---Wtt 2018) specifies five types of corporate services that require a license: ``Being a director/partner of a legal entity/company; providing an address or postal address for an object company and performing additional activities, such as record-keeping or preparing and filing tax returns; selling or intermediating in the sale of legal entities; acting as a trustee; and the provision of a conduit company.'' \cite{dnbWhatAreTrust2021}. The provision of these services without a license is considered illegal.\footnote{Corporate service providers can offer other services without a license (e.g. corporate tax advice). In this paper we define CSPs as entities (individuals or corporations) providing services that require a license under the Wtt 2018. Companies providing these services are known in Dutch as ``trustkantoren'', a term with no direct translation.}  Here, we focus on the provision of nominee directors---CSPs managing companies on behalf of a client---and refer to unlicensed CSPs as illegal CSPs.

%Gap
Over the last decade, the Dutch government and the Dutch central bank have gradually been tightening the regulation and supervision of the CSP industry. Parallel to this development, the total number of licensed CSPs have decreased by 46\%, from 192 in October 2013 to 103 in January 2020, while the costs of supervision---which are borne by CSPs and involve inspecting CSPs and examining signals of illegal activities (see Section \ref{app:supervision})---increased by 216--640\%. Anecdotal evidence suggests that this decline in licensed CSPs has been at least partially offset by CSPs that operate without a license \cite{loggerNederlandseFlexkantorenBloeit2018}. The size of this illegal CSP industry is, however, unclear. Over one million persons act as directors of companies in the Netherlands, and only a few hundreds or thousands provide illegal corporate services. Illegal CSPs can be seen as the proverbial needle in a haystack.

\begin{figure}[h!]
    \includegraphics[width=0.5\textwidth]{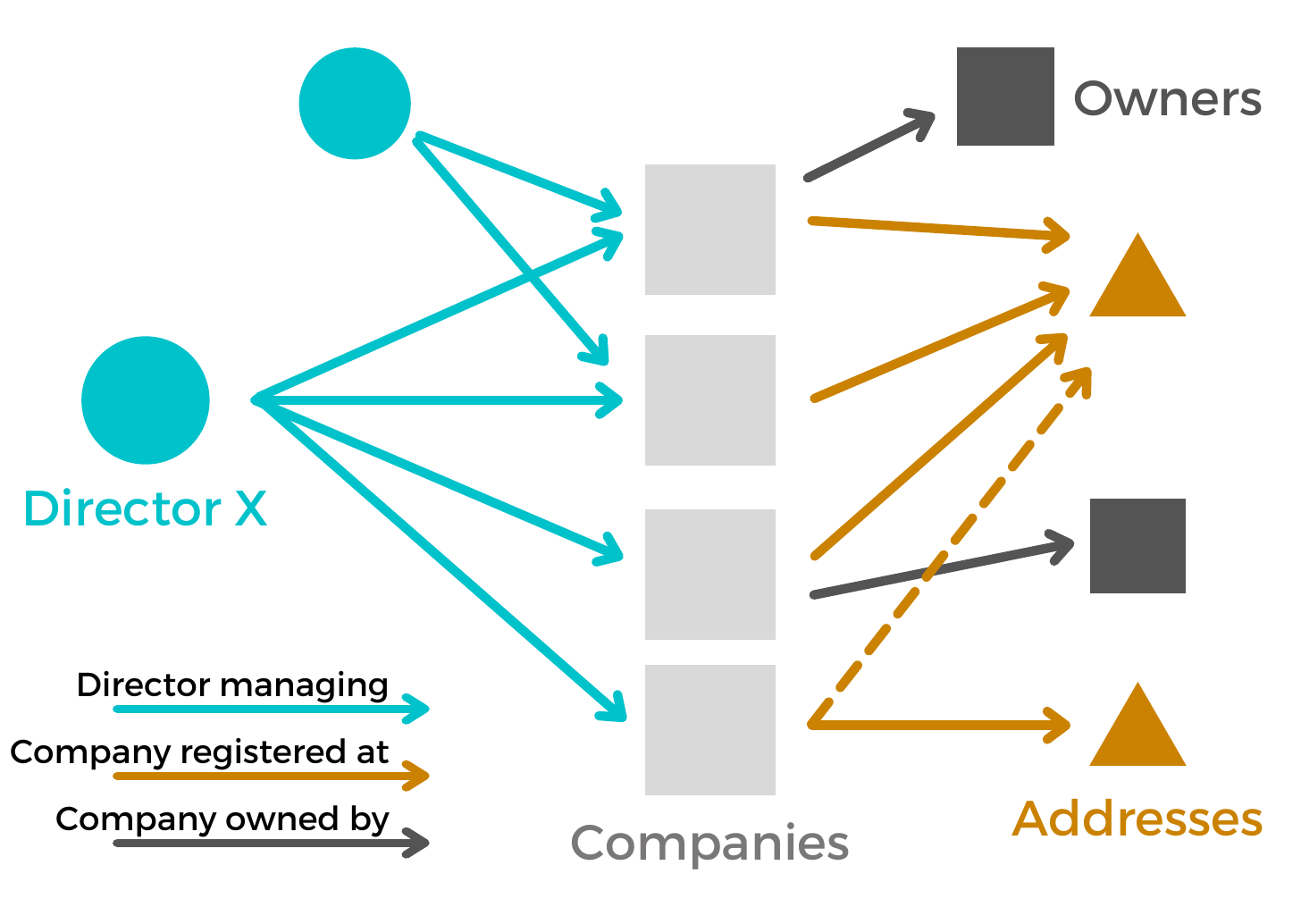}
    \caption{Schematic representation of the two-step ego-network of director X. Director X (middle turquoise circle) manages four companies (gray squares),  registered at specific addresses (brown triangles). The companies can be owned by other companies or individuals (black squares), can have additional postal addresses (dashed brown line) and additional directors (top turquoise circle).\vspace{1cm}}
    \label{fig:network_scheme}
\end{figure}

%Results
In this paper, we develop a classifier to estimate the number of the illegal (i.e., non-licensed) CSPs in the Netherlands. This classifier is based on the network formed by Dutch corporations, their directors, the addresses at which they are registered, and their owners.
Directors with a similar position in the network are expected to have a similar function. For instance, directors managing a large number of companies, owned by foreign individuals, and registered in the same address are likely to be CSPs. By comparing subnetworks of licensed CSPs (see Figure \ref{fig:network_scheme} for a schematic depiction of one such subnetwork) to all other directors in the network, we find a subset of illegal directors that are ``sufficiently similar'' to licensed CSP directors, such that we deem them of high risk of being part of illegal CSP networks.
We then manually annotate 200 of them using different online sources (e.g., personal websites, corporate websites, LinkedIn, trial cases) to estimate that in 2019 there were 402 CSPs (95\% confidence interval: 212--668) with a very high risk of offering illegal\footnote{Our estimate is based on the team's expertise and does not constitute a formal regulatory or judicial verdict regarding the factual ``illegality'' of the provided services. Throughout the paper we use ``illegal services'' instead of ``very high risk of offering illegal services''.} directorship services to 2,414 companies (95\% confidence interval: 1,274--4,012). This corresponds to 31\% of all CSPs and 19\% of all companies managed by CSPs. These numbers increase to 38--51\% and 23--27\% respectively when we extrapolate our results to include small illegal CSPs---those managing one or two companies. 
%Our results allow us to understand the reach of the trust supervision in the Netherlands. 
Our network-based approach could be applied as a red-flag system to monitor the CSP industry and reduce illegal activities. Our approach could also be adapted in a variety of networks where a fraction of the nodes is flagged and the goal is to find similar cases among the unflagged nodes---for example in customer due diligence, the process used by financial institutions to identify, screen and monitor their customers to prevent financial crime.

The remainder of this paper is organized as follows.
Section \ref{sec:csp_ndl} briefly describes how CSPs are regulated in the Netherlands. 
Section \ref{sec:data_methods} describes the datasets used and the cleaning and data augmentation process, the engineering of features based on network characteristics, and the algorithms used to predict directorship services. Section \ref{sec:results} presents the results of the analysis. Finally, section \ref{sec:conclusion} concludes by summarizing our results and suggesting some of the policy and future research implications of these results.

\section{Corporate service providers in the Netherlands (``\textit{trustkantoren}'')  }  \label{sec:csp_ndl}
%CSPs in the Netherlands
Corporate service providers facilitate the incorporation and management of companies. In the Netherlands, this industry primarily caters to foreign direct investment (FDI) in and out of the country---FDI stocks in the Netherlands rank second globally, only after the United States \cite{unctadWorldInvestmentReport2020}. 
The Netherlands is renowned for its business orientation---for example attractive fiscal policy, good infrastructure, skilled workforce, access to deep and developed capital markets, low corruption, and presence of a strong CSP industry.
Setting up and maintaining corporate entities in the Netherlands, and thus gaining access to Dutch legislation, is relatively inexpensive and provides access to a wide range of advantageous tax and investor protection regulations.
Consequently, the Netherlands is one of the largest conduit countries in the world \cite{garcia-bernardoUncoveringOffshoreFinancial2017}, acting as an intermediate destination for global corporate financial flows.
A recent report by the ``Commissie Doorstroomvennootschappen'' found that, as in 2019, financial flows through conduit companies in the Netherlands amounted to 4,500 billion. Around 65\% of those companies in the Netherlands were managed by CSPs \cite{commissiedoorstroomvennootschappenOpWegNaar2021}.

%Illegal CSPs in the Netherlands
In the Netherlands, the CSP sector is regulated by the Act on the Supervision of Trust Offices 2018 (``Wet toezicht trustkantoren'', Wtt 2018), which establishes that the provision of certain corporate services, such as becoming a nominee director for a company, require a license and the obligation to know and monitor the client---thereby reducing the risk of accepting clients involved in money laundering, bribery, or terrorism financing. However, setting up a company on behalf of somebody else, or acting as a nominee director can easily be arranged at the chamber of commerce without the need of a CSP license, which provides opportunities for an illegal market of corporate services. Irrespectively of the activity taking place in the company, becoming a nominee director without a license constitutes an illegal activity. 

Over the last decades, Dutch policy makers have gradually strengthened the regulation of the Dutch CSP industry. The Wtt 2018 imposed several new requirements on CSPs, such as mandatory incorporation as a juridical person, a minimum number of board members, as well as further requirements on customer due diligence and auditing functions. This increase in legal requirements has been matched by increased regulatory supervision. The conjunction of these developments has resulted in higher costs for regulated CSPs in two ways. First, CSPs have had to increase compliance expenditures in order to meet the more stringent requirements in the Wtt 2018. Second, within the Dutch regulatory context, the costs of supervision are passed on to the market participants subject to such supervision (Section \ref{app:supervision}). Taken together, the higher costs for CSPs and the increased legal requirements have contributed to a decline in the number of registered CSPs by 46\%, from 192 in October 2013 to 103 in January 2020. On a per license basis, this decrease in the number of registered CSPs further raises the costs of supervision for the remaining CSPs given that not all costs of supervision scale with the number of CSPs.  

The increase in the regulatory burden for CSPs has strengthened the incentives for market participants to offer services just outside of the scope of the Wtt 2018 or in non-compliance with the Wtt 2018. Indeed, anecdotal evidence suggests that the decrease in licensed CSPs has at least been partially offset by unlicensed service providers \cite{loggerNederlandseFlexkantorenBloeit2018}. From a financial crime perspective, this may increase the risks. Firms outside of the scope of the Wtt 2018 are subject to less stringent supervision given that only the Dutch Act on the Prevention of Money Laundering and Financing on Terrorism (``Wet ter voorkoming van Witwassen en financieren van terrorisme'', Wwft) applies to this group, instead of both the Wwft and the Wtt 2018. 

\section{Data and methods} \label{sec:data_methods}
\begin{figure}[b!]

    \includegraphics[width=0.9\textwidth]{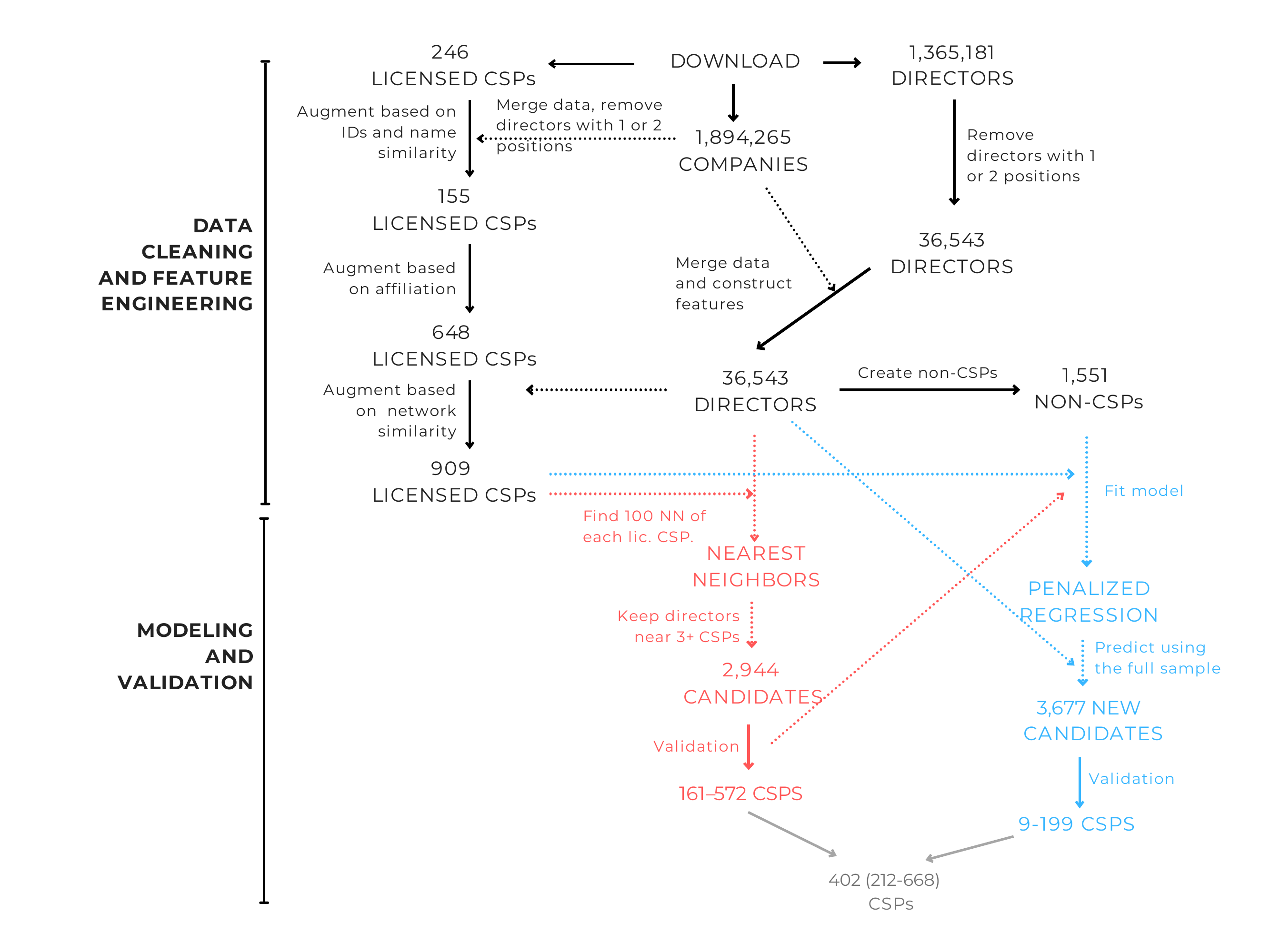}
    \caption{Flowchart with our approach, consisting of two phases. (i) The data cleaning and feature engineering phase (top part, in black). We started by downloading the datasets on licensed CSPs, companies, and directors. Directors (right branch) with one or two positions are excluded, and the remaining 36,543  directors are merged with company data to construct the network features. Licensed CSPs (left branch) are augmented using company data to reach a final set of 909 licensed CSPs. (ii) The modeling and validation phase (bottom part, in color). First, we use the nearest neighbors algorithm to find similar directors to the 909 licensed CSPs (red branch). We kept all directors that were within the 100 closest directors to at least three licensed CSPs. We manually validated 100 of them to estimate the size of the illegal CSP industry at 161--572 entities.  Second, we conducted a validation test using penalized logistic regression (blue branch). We found 3,677 new potential candidates and manually validated 100 of them to estimate that the first approach missed 9--199 illegal CSPs. Taken together, we estimate the size of the illegal CSP industry at 402 entities (95\% confidence interval 212--668).  Solid arrows indicate a transformation or creation of a dataset. Dashed arrows indicate inputs.}
    \label{fig:flowchart}
\end{figure}

We identify directors at high risk of providing corporate services without a license based on their network similarity with licensed CSPs. For example, both will tend to serve as directors for companies that are registered at the same address and that are owned by foreign owners. In this section, we detail the data sources that we use (Section \ref{sec:data}), the network-based features that we create to fit our model (Section \ref{sec:methods:network}), and the classification algorithms that we apply (Section \ref{sec:models}). Our entire approach is summarized in Figure \ref{fig:flowchart}.

\subsection{Data and feature construction} \label{sec:data}
We obtained our main dataset from the chamber of commerce (\url{https://kvk.nl}) through  the Orbis database (\url{https://bvdinfo.com})---a commonly used corporate information provider (Section \ref{sec:data:orbis}). We downloaded the full dataset, 1,894,265 companies and their 1,365,181 directors---36,543 of them holding positions in at least three companies. These data allow us to understand the characteristics of the companies managed by those directors (Section \ref{sec:methods:network}). We were able to match 909 of these directors to the list of CSPs registered at the central bank (Section \ref{sec:trust}), and created classification models (Section \ref{sec:models}) to find similar directors that may be providing corporate services. 

\subsubsection{Corporate and directorship data} \label{sec:data:orbis}

For each of the 1,894,265 companies (identified by name and company ID) and year, we obtained the following fields (exact variable names can be found in Appendix \ref{app:orbis}):
\begin{itemize}[leftmargin=*]
\item Current and previous addresses: Street, street number, postcode, city, type of address (office or postal). Previous addresses were not readily available, but had to be obtained from the Orbis variable ``Legal events - Description'', which required dividing the string into the different address fields (street, number and postcode) using regular expressions (Section \ref{app:regex}). The combination of a postcode and number uniquely identifies addresses in the Netherlands.
\item Current and previous directors: Director ID (created by Orbis), company ID (for the cases when the director is a corporate entity and not an individual), director name, director title, status (current or previous). We conceptualize ``directors'' as a collective category of relevant decision makers and authorized representatives within an entity (e.g. statutory directors, proxy holders, etc.). We obtained 1,365,181 directors. Since we rely only on publicly available data, distinguishing between a CSP providing services to only one company and a freelancer would be extremely difficult and the results would be impacted by noise. To create a more accurate prediction model, we filtered out all directors with one or two positions to obtain a list of 36,543 directors. In section \ref{sec:extrapolation} we extrapolate our results to include small directors, and detail the potential influence of this choice.
\item Financial and industry information: NACE rev. 2 sector of the company, assets, turnover, employees, profits.
\item Ownership: Location of the global ultimate owner (GUO) of the company.
\end{itemize}

\subsubsection{Information on licensed CSPs} \label{sec:trust}
The Dutch Central Bank (DNB), responsible for the supervision of CSPs, publishes a list of all licensed CSPs \cite{dnbRegisterTrustOffices2021}. We collected information on 246 CSPs, registered in 139 unique addresses with 239 unique company names. Since the postcodes of the CSPs' addresses provided by the DNB were not readily available, we used the google geocoder service \cite{OverviewGeocodingAPI2021} to obtain them.

The next step is to augment the data in order to capture the entire legal CSP industry (Fig. \ref{fig:flowchart}, left side on ``licensed CSPs''). There are two types of directors that we needed to include: \vspace{-0.8em}

\hspace{3em}  (i) branches of CSPs (that do not need to register independently). We added these by matching the national identification number (KvK nummer) of the licensed CSPs to the Orbis database, and obtaining all branches linked to the same IDs. We were able to match 195 CSPs, which allowed us to increase the number of addresses associated to CSPs to 409 (270 extra) and the number of company names to 244 (5 extra). Then, we matched our list of 244 unique names to the list of directors obtained from Orbis and were able to match 220 of them to unique director IDs (which identify the director in Orbis). Finally, we matched the 244 company names to the list of directors downloaded from Orbis by name similarity. To do so, we split the directors names (which can be either individuals or companies) into trigrams and normalized their count using term frequency-inverse document freqency (TF-IDF). We added all directors with a cosine similarity above 90\% to the list of licensed CSPs, which added 9 extra entities (for a total of 229). We chose the 90\% threshold via iterative examination of the results---e.g. Vistra Management Services B.V. and Vita Management Services B.V. have a similarity of 80\% and are different companies, Sempter Fidelis B.V. and Sempter Fidelis Beheer B.V. have a similarity of 92\% and correspond to the same corporate group). Of the 229 corporate directors matched Orbis, 155 entities provided services to three of more companies. \vspace{-0.8em}

\hspace{3em}  (ii) the employees of the CSPs. We identify these in two ways. (iia) We included the 493 directors of the 229 corporate directors matched in the previous step. We assumed that these directors would only provide corporate services through their associated CSP (i.e., they would not provide illegal services). This increased our list of licensed CSPs to 648 entities. (iib) We identified 261 extra directors based on the characteristics of the companies they manage. We use two overlapping criteria: directors with over 25\% of their managed companies registered at the office or postal address of a licensed CSPs (using the DNB list) and where over 20\% of their companies have another director from a licensed CSP (257 cases); directors with over 50\% of their managed companies registered at the office or postal address of a licensed CSPs (using the augmented list, i.e., after adding the addresses found in Orbis) and where over 50\% of its companies have another director from a licensed CSP (5 extra cases). We tested our approach using manual annotation of 100 directors (section \ref{sec:manual_annotation}), in which 15 out of 15 directors in the sample found in step iia were in fact working for a licensed CSP, and 3 out of 5 directors of step iib were working for a licensed CSP (we were unable determine one, and one was a Chinese official linked to a licensed CSP, but not providing corporate services). Moreover, we tested the effect of not including the directors of step iib: 86\% (3,277 out of 3,830) of the potential CSPs detected by our algorithm were still detected if the directors included in step iib were excluded.
The final list of licensed CSPs managing three or more companies consists of 909 entities. In total, these licensed CSPs hold 17,924 directorships in 6,913 independent companies: 6,992 directorships held by the 155 CSPs of step (i), 8,700 directorships held by the 493 CSPs of step (iia), and 2,223 directorships held by the 262 directors of step (iib).

\subsubsection{Creating a sample of directors not providing corporate services} \label{sec:noncsp}
Directors (e.g. CEOs) are typically hired based on their industry expertise for the strategic planning and oversight of a company. It is only when they manage the company on behalf of a client that they provide corporate services and a license is required. 
%Nominees directors represent primarily the interests of the business owner instead of promoting the general success of the company. 
In this paper, we use penalized logistic regression to validate the results of our main method based on the nearest neighbors algorithm. Logistic regression (which will be explained in Section \ref{sec:logit}) requires examples of CSPs (for which we use licensed CSPs) and examples of non-CSPs. 
We use all corporate directors (i.e., non-individuals) in productive sectors as examples of directors \textit{not} providing corporate services. We define productive sectors as all sectors in the NACE rev. 2 classification except 64--66 (Financial and insurance activities), 69 (Legal and accounting activities), 70 (Activities of head offices; management consultancy activities) and 82 (Office administrative, office support and other business support activities).

%% We only used this for the domiciliation analysis, not the directors
% \subsubsection{Land registry (kadaster}

% An indication of an address being used to provide domiciliation services (the provision of registered addresses is consider a trust service if provided alongside any other administrative or tax service) is the ratio of square meters to number of companies registered at that address. Ratios below $25 m^2/comp$ are a strong indication that companies do not carry physical activities at the place. In order to obtain this information, we used the data from the land registry \url{https://www.kadaster.nl/zakelijk/registraties/basisregistraties/bag}. 

% For each of the 7,858,497 addresses (identified by their postcode and street number), we calculated the square meters, and the type of function (e.g. office, living, store, industry). Sub-addresses (e.g., Street A 34-A) were summed together (e.g., to Street A 34) since Orbis does not accurately parse this information. This is a conservative approach since the $m^2$ per company will tend to be larger.

\subsubsection{Offshore leaks database}
Finally, we collected the offshore leaks database, provided by the International Consortium of Investigative Journalism \cite{icijHowDownloadThis2020}. This database contains a list of 504,851 addresses, 861,576 company names, 832,468 officers, and 13,203 intermediaries (e.g. tax officers or corporate service providers) involved in the Bahamas leaks, Offshore leaks, and Paradise leaks. We use these data to flag all addresses, company names or director names present in both the offshore leaks and the Orbs data. Flags are created when the address/company/director in the offshore leaks database is contained in the address/company/director name in Orbis. While being present in the offshore leaks does not imply illegal activities, it is a sign of potential illegal activities \cite{odonovanValueOffshoreSecrets2019}. 

\subsection{Network approach to construct features}\label{sec:methods:network}
We use the data sources described in section \ref{sec:data} to create a network of directors, companies, addresses and owners (Fig. \ref{fig:network_scheme}).
The network is based on the following relations: directors---companies; companies---addresses (postal and office); companies---owners.

We then find potential illegal CSPs based on their similarity with licensed CSPs, where their similarity is computed based on their position within the network of directors, companies and addresses, and the characteristics of such entities. Leveraging the information contained in the relationships between network entities (companies, directors, etc) has often been proposed or used to detect economic crime \cite{sparrow1991application,klerks2004network,kertesz2021complexity, wachs2021corruption}. We measure similarity  based on 48 indicators at the director level. The indicators (Table \ref{tab:variables}) are created based on the expertise of the team (the authors and the consultants detailed in the acknowledgments) and information from semi-structured interviews and workshops with experts and industry representatives (see Section \ref{app:interviews} for more information on the interview process). They are calculated based on the ego network of each director at depth two---i.e., director X, the companies that are managed by director X, and the addresses, owners and other directors of such companies (see Figure \ref{fig:network_scheme}). We display the indicators in Table \ref{tab:variables}, organized according to the entities in Figure \ref{fig:network_scheme} that relates to them.  
The indicators mainly look at characteristics of the director (e.g. is the director an individual or a corporate entity?), characteristics of the combination of directors and companies (e.g., is the name of the director similar to the name of the companies?), characteristics of addresses (e.g. how many companies are registered at a specific address?), characteristics of the companies (how many companies does a director manage?), characteristics of the owners (e.g. foreign vs domestic owners), and characteristics of both owners and companies (e.g., number of independent companies managed.  Independent companies correspond to the ultimate owner of the companies managed, or in the case where no owner is recorded, the company itself). We do not perform feature selection since assessing the algorithm performance would require labeled data. We, however, checked that our results were robust to feature selection by comparing the agreement of our algorithm with the results of the algorithm trained with a subset of the features (see Appendix \ref{sec:robust_var_selection}). We detail in Appendix \ref{sec:variables_detail} how each indicator is created and the expected impact on the likelihood of a director providing corporate services requiring a license.

\begin{table}[]
\begin{tabular}{p{8.5cm}p{7.5cm}}
\textbf{\uline{Characteristics   of directors}}                                            & \textbf{\uline{Characteristics of the companies}}          \\
Is the director an  individual   or a corporate entity?                   & Any company in a CSP-related sector?      \\
Director's   name contains   a corporate keyword                          & Log  number of companies                  \\
Corporate director in a CSP-related   sector                              & String   similarity between company names \\
Director appears in offshore leaks                                        & \%   of companies in finance              \\
                                                                          & \%   of holding companies (NACE 6420)     \\
\textbf{\uline{Characteristics of both   directors and companies}}                         & \%   of real estate companies             \\
Corporate  director: shared   directors with companies                    & \%   of top holdings                      \\
Name similarity between companes and   director                           & \%   of administrative companies          \\
\%   of companies with the   title ``director''  & \%   of companies with unknown sector     \\
\%   of companies with the most frequent title     & \%   of companies in retail/wholesale     \\
Number     of shared directors between the companies                      & \%   of companies in construction         \\
Average  number of previous   directors per company                       & \%   of companies of top sector           \\
\%   of companies with a   previous licensed CSP                          & \%   of BVs                               \\
                                                                          & \%   of Foundations                       \\
\textbf{\uline{Characteristics of the   addresses}}                                        & \%   of VOFs                              \\
Log   \# of companies in   the top office address                         & \%   of Cooperatives                      \\
Log   \# of companies in   the top poastal address                         & \%   of CVs                               \\
\%   of companies in the   top office address                             & \%   of companies with the top legal form \\
\%   of companies in the   top postal address                             & Number   of companies in offshore leaks   \\
\%   of companies   previously in an address of CSP (augmented)           &                                           \\
\%   of companies   previously in an address of CSP                       &                                           \\
Number  of office addresses in   offshore leaks                           & \textbf{\uline{Characteristics of the owners }}            \\
Number  of postal addresses in   offshore leaks                           & Number of directors that are also owners \\
Average   number of   previous address                                    & \%   of directors that are also owners    \\
                                                                          & \%   of companies with unknown owner      \\
\textbf{\uline{Characteristics of both owners  and companies}}                            & \%   of companies with domestic owner     \\
Log  number of independent   companies                                    & \%   of companies with foreign owner      \\
Number of directors per independent   company                             & \%   of companies with owner in an OFC    \\
Number of companies per independent   company                             &                              \vspace{0.5cm}            
\end{tabular}
\caption{Features constructed according to the category: features related to characteristics of the director, of the addresses, of the companies, of the owners, or of combinations. A full description of each variable can be found in Appendix \ref{sec:variables_detail}.}
\label{tab:variables}
\end{table}

\subsection{Modeling and validation} \label{sec:models}
After creating the network features for the 36,543 directors we used a nearest neighbors algorithm to find directors with similar characteristics to licensed CSPs. 
This algorithm allows us to flag potential CSPs without relying on labeled data on non-CSPs---as it would be the case with model-based classifiers.
We used an inclusive approach, aimed at capturing the vast majority of potential CSPs. We then manually annotated a subsample of 100 entities potential CSPs to estimate the size of the illegal CSP sector. To validate the results, we needed to understand if the nearest neighbors algorithm was actually able to detect the vast majority of CSPs. To check this, we used a penalized logistic regression to flag a new sample of 3,691 directors not previously found by the nearest neighbors algorithm. We manually annotated a subsample of 100 directors to confirm that the nearest neighbors algorithm had found the majority of illegal CSPs. For both algorithms we used the implementation in the Python library \textit{scikit-learn} \cite{pedregosa2011scikit}.

\subsubsection{Nearest neighbors} \label{sec:model:nn}
\begin{figure}[h!]

    \includegraphics[width=\textwidth]{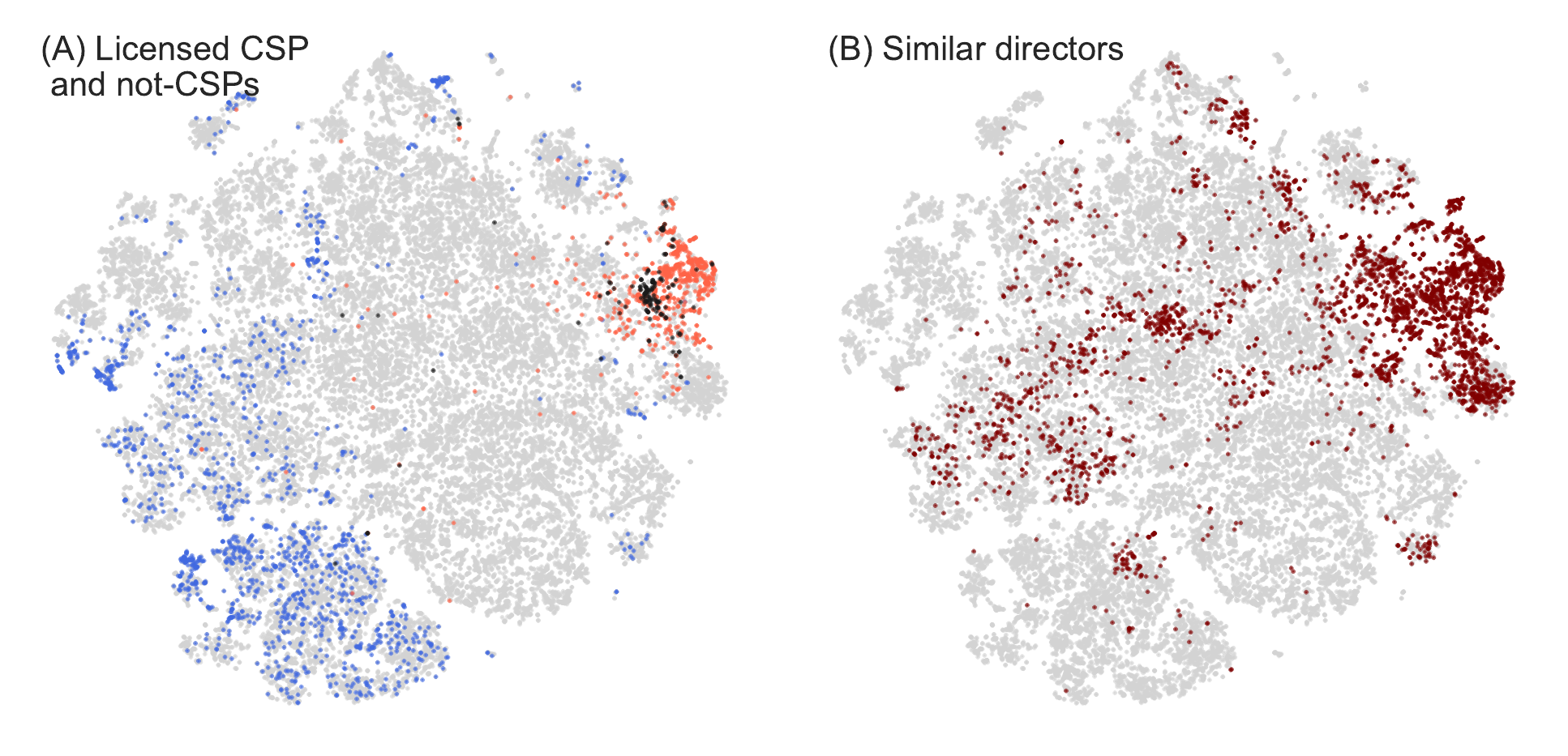}
    \caption{Visualization of the feature space using t-distributed stochastic neighbor embedding (t-SNE). Each point represents one director. Points that are near each other in the feature space appear close in the visualization. (A) Licensed CSPs are visualized in black (original list from the Dutch central bank) and red (augmented list as detailed in section \ref{sec:trust}). Non-CSPs (see section \ref{sec:noncsp}) are visualized in blue. (B) The nearest neighbors to the licensed CSPs are visualized in maroon. Only those points that are close to at least three CSPs are displayed (i.e., they are within the 100 closest neighbors of three CSPs). }
    \label{fig:tsne}
\end{figure}

Starting from our list of 909 licensed CSPs (directors), we obtained the 100 closest neighbors to each one in the standardized feature space (Fig. \ref{fig:tsne}).
These 100 neighbors can themselves be licensed CSPs.
In order to find the nearest neighbors we need to define the distance between two directors. We use the Eucledian distance  $\sqrt{\sum_i^{i=48} (x_i^1 - x_i^2)}$, where $x_i^1$ and $x_i^2$ are the standardized value of feature $x_i$ for the two directors. We used the implementation of KD-Tree algorithm in the Python library \textit{scikit-learn} to  increase the performance of the algorithm.
We flagged as potential CSPs only those directors that were selected over three times (3,830 potential directors). This threshold was selected to retrieve a large number of neighbors and capture the majority of CSPs (see Fig. \ref{fig:var_selection}). A higher threshold (obtaining a smaller sample) would have also been valid, but this was unknown to us until we manually annotated a sample of them. Of the 3,830 potential directors, 886 were licensed directors (i.e., 97.5\% (886 out of 909) of the licensed directors were included. Some licensed directors were not included because they were not within the 100 closest neighbors of three other licensed directors) and only 23 were directors not providing corporate services (2.5\% false positive rate in that subsample). We are left with 2,944 potential candidates (see also the flowchart in Figure \ref{fig:flowchart}).
We use such an inclusive approach in order to increase the recall of the model---i.e., reduce false negatives at the expense of increasing false positives (Fig. \ref{fig:app_roc}). 
Our approach is tailored to our research question---estimate the size of the illegal CSP sector. Given that we are only annotating a subsample of the directors found by the algorithm we wanted to reduce the probability of our algorithm not finding CSPs (false negatives). 
The presence of false negatives would lead to a biased estimation of the size of the illegal CSP sector, while the presence of false positives would only increase the confidence interval of our estimate.
Different research questions, such as finding illegal CSPs for police investigations, may require to balance false negatives and positives differently. 
For example, if the cost associated to the audit of a potential illegal CSP is high we could use a less inclusive approach (a higher threshold in our algorithm) and reduce the number of false positives.

Figure \ref{fig:tsne} shows a projection of the feature space into two dimensions using t-distributed stochastic neighbor embedding (t-SNE) \cite{van2008visualizing}. This dimensionality reduction technique maps similar observations in the 48-dimensional space nearby in a 2-dimensional space, and dissimilar observations far apart. We find that licensed CSPs (in black and red) and non-CSPs (in blue) occupy very different and compact parts of the space (Fig. \ref{fig:tsne}A), which indicates that our features are able to capture the characteristics of CSPs. Figure \ref{fig:tsne}B shows the location of the potential candidates to be CSPs. While the majority of licensed CSPs cluster in the ``top right corner'' in Figure \ref{fig:tsne}A, some licensed CSPs spread over the entire space (notice the dispersed red dots). As a result, the algorithm finds most directors in the ``top right corner'', but also some throughout the space (Fig. \ref{fig:tsne}B).

\subsubsection{Penalized logistic regression} \label{sec:logit}
The nearest neighbors method ensured that 97.5\% of licensed CSPs were captured. The subsequent manual validation showed that illegal CSPs are likely to exist, and that our method is able to retrieve at least some of them. Even while we used an inclusive approach to increase the recall (minimize false negatives) of our approach, we still have no information about the exact recall of our model. That is, we cannot establish to what extent we captured all CSPs. In order to measure the recall we could manually annotate a random sample of the entities labeled as not providing CSPs that were not found previously by the Nearest Neighbors algorithm. Given the low prevalence of illegal CSPs in the network, however, this would require an impractically large sample size. 

In order to find a subsample that we can manually annotate, we fit a L2-penalized logistic model to our data of 36,543 directors. 
A non-penalized logistic model would estimate the probability of being a CSP as $p(CSP) = 1/\left(1+e^{-\sum_i^{i=49}(W_i x_i)}\right)$, where $x_i$ corresponds to feature $i$ and $W_i$ are the associated coefficients.\footnote{Note that there are 49 coefficients since we include the intercept as the coefficient of a feature consisting of ones.} A L2-penalized model adds an extra term to the cost function (which is used to train the model) equal to $\lambda \sum_i^{i=49}(W_i^2)$. The cost function  thus depends on both the difference between the predictions and the real values and the value of the weights. This shrinks the estimates of the weights and prevents overfit arising from high-dimensionality. The logistic model requires positive examples (examples of CSPs) and negative examples (examples of non-CSPs) to be trained.
For the positive examples we use the 909 licensed CSPs and 8 illegal CSPs found via manual annotation of the results of the nearest neighbors algorithm. We use non-CSPs (section \ref{sec:noncsp}) as our negative examples. The optimal regularization strength was estimated through cross-validation. Similarly to the nearest neighbors method case, we aimed at capturing most potential illegal CSPs, at the expense of increasing the number of false positives. In the nearest neighbors method, this was done by increasing the number of neighbors. In the logistic regression, this was done by retrieving all directors with a predicted probability of being CSPs above 1\%.

\subsubsection{Manual annotation}  \label{sec:manual_annotation}
In order to understand the precision---the fraction of entities labeled as CSP who are actually CSPs---we manually coded 100 of the entities detected by each algorithm. This was done by a research assistant using a code book created by the authors (see Appendix \ref{sec:manual_annot}), a knowledge graph of the network (see Appendix \ref{sec:neo4j}), as well as information on the companies managed by the director.
In particular, we took into consideration the postal, office and postbus addresses of the companies (the number of companies in that address, the presence of the address in the offshore leaks, and the presence of licensed CSPs in the address), the country of the company owner, the number of directors of the companies, and the sector and type of legal entity of the company.
Taking all the information together, we evaluated the probability of providing corporate services in a 5 point scale, where directors labeled as 4 were considered likely to provide corporate services and 5 were considered almost certain to provide corporate services. The coding was then reviewed independently by JW and JGB (the authors), and adjusted as necessary.
Directors labeled as either 4 or 5 are considered CSPs. Cases were the two coders disagreed (one coder evaluating it as a 1, 2, or 3 and the other as a 3, 4, or 5) were marked as unknown (i.e., 3).

Given that the number of illegal CSPs in the 100 entities labeled follows a binomial distribution (with probability of being illegal $\theta$), we can analytically estimate the binomial proportion $\theta$ using Bayesian inference (see e.g., chapter 8 of \cite{bolstad2016introduction}). We use a uniform prior---i.e., $\mathcal{B}eta(1, 1)$---since we have no knowledge a priory of the performance of our algorithm. The posterior distribution of $\theta$ is given by 
$P(\theta \mid TP, FP) = \mathcal{B}eta(TP+1, FP+1)$,
where \textit{TP} is the number true positives and \textit{FP} is the number false positives (total candidates - true positives).

The distribution of the expected number of illegal CSP in the full sample is given by the product of the total number of candidates flagged by the algorithm, and the distribution of $\theta$, $P(\theta \mid TP, FP) $. We calculate confidence intervals as the 95\% confidence interval of the posterior distribution and use the median as our point estimate. 

For the nearest neighbors approach, we label a random subsample of 100 directors found by the algorithm. For the logistic regression approach, since we were interested in understanding how many illegal CSPs remained undetected by the nearest neighbors approach, we labeled 100 directors not previously found by the nearest neighbors approach and not licensed.

We estimated the \textit{total} size of the illegal sector combining both algorithms. For this, we take 1,000,000 samples from the distribution of the number of illegal CSPs calculated for each algorithm and sum the results. We end up with 1,000,000 values representing the expected number of illegal CSPs in the population. Similarly to the previous case, we calculate the 95\% confidence interval of the distribution and use the median as our point estimate.

\section{Results and Discussion} \label{sec:results}
\subsection{Size of the illegal and legal sector} %%Javier: I want to adapt some numbers, in the report we used medians but I think we should use means.

\begin{figure}[h!]

    \includegraphics[width=\textwidth]{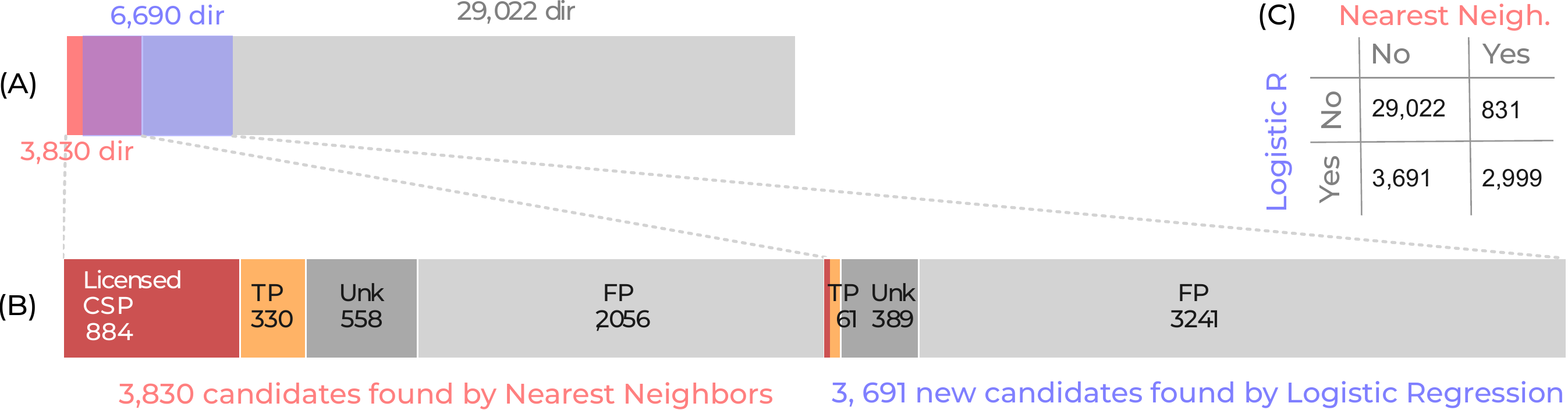}
    \caption{Number of directors flagged and validated in our approach.  (A) The nearest neighbor approach (red) identifies 3,830 directors as potential CSPs, while the logistic regression (blue) identifies 6,690 (3,691 new ones). (B) Amongst the directors flagged by the nearest neighbors approach, 886 correspond to licensed CSPs, 330 to illegal CSPs (TP), 2,056 to non-CSPs (FP), and we were not able to determine the status of 558 of them (Unk). Amongst the new directors flagged by the logistic regression approach, 12 correspond to licensed CSPs, 61 to illegal CSPs (TP), 3,241 to non-CSPs (FP), and we were not able to determine the status of 389 of them (Unk). The estimates of false positives, true positives and unknowns were obtained using the Bayes rule with a uniform prior and a binomial likelihood. The median of the posterior distribution is displayed. (C) Confusion matrix with the overlap between the directors flagged by both algorithms.}
    \label{fig:summ_bayes}
\end{figure}

Figure \ref{fig:summ_bayes} summarizes the application of our estimation methodology (depicted in Fig. \ref{fig:flowchart}) at the director level. Out of the 36,543 directors, our nearest neighbors algorithm flags 3,830 directors. Of those, 886 correspond to licensed CSPs, and 2,944 are potential illegal CSPs We manually labeled a random sample of 100 of them to find that 11\% are at high risk of providing illegal services. Extrapolated at the full population, this involves 161--572 directors providing illegal services (95\% Bayesian confidence interval using a uniform prior).

Our algorithm greatly reduces the need to manually label observations. If we were to draw a random sample from the population of 36,543 directors, the sample size required to estimate the size of the illegal CSP sector with a similar accuracy would need to be 11 times larger. Assuming that there actually exist 330 illegal CSPs in the Netherlands, annotating a subsample of 1,100 directors out of 35,634 directors (finding approximately 10 illegal CSPs), we would get a confidence interval comparable to the one found with our method (178--588 vs 161---572).\footnote{The shift in the center of the confidence interval is due to the influence of the uniform prior used to calculate confidence intervals.} As additional advantages to our approach, our results can be extrapolated to include small CSPs (detailed in Section \ref{sec:results:small_entities}), and our classifier could be used as a red-flag system to monitor the CSP sector.

%How many directors

We then explored if the nearest neighbors algorithm was able to find all CSPs. For this, we fitted a L2-penalized logistic regression model (Section \ref{sec:models}), which allowed us to flag as potential CSPs 3,677 directors not detected by the nearest neighbors approach and not licensed. We manually label a random sample of 100 of them to find that 1\% are at high risk of providing illegal services. This implies that the nearest neighbor algorithm missed 9--199 illegal CSPs. Our two methods flag a similar set of directors, as reflected in the overlap in Figure \ref{fig:summ_bayes}A and the confusion matrix in Figure \ref{fig:summ_bayes}C. 2,999 directors were flagged by both algorithms. Of those, 863 correspond to licensed CSPs (95\% of the total number of licensed CSPs). 

Combining the results of both algorithms (nearest neighbors and logistic regression), we estimate the size of the illegal CSP sector at 402 directors, with a 95\% confidence interval of 212--668. Relative to the 909 identified licensed directors, this implies that the ``illegal'' market share is 31\% (19--42\%). Given that we exclude cases where we could not identify if a director was a CSP or not (dark gray bars in Figure \ref{fig:summ_bayes}B), this market share can be considered as a lower bound.

\begin{figure}[h!]
    \includegraphics[width=0.6\textwidth]{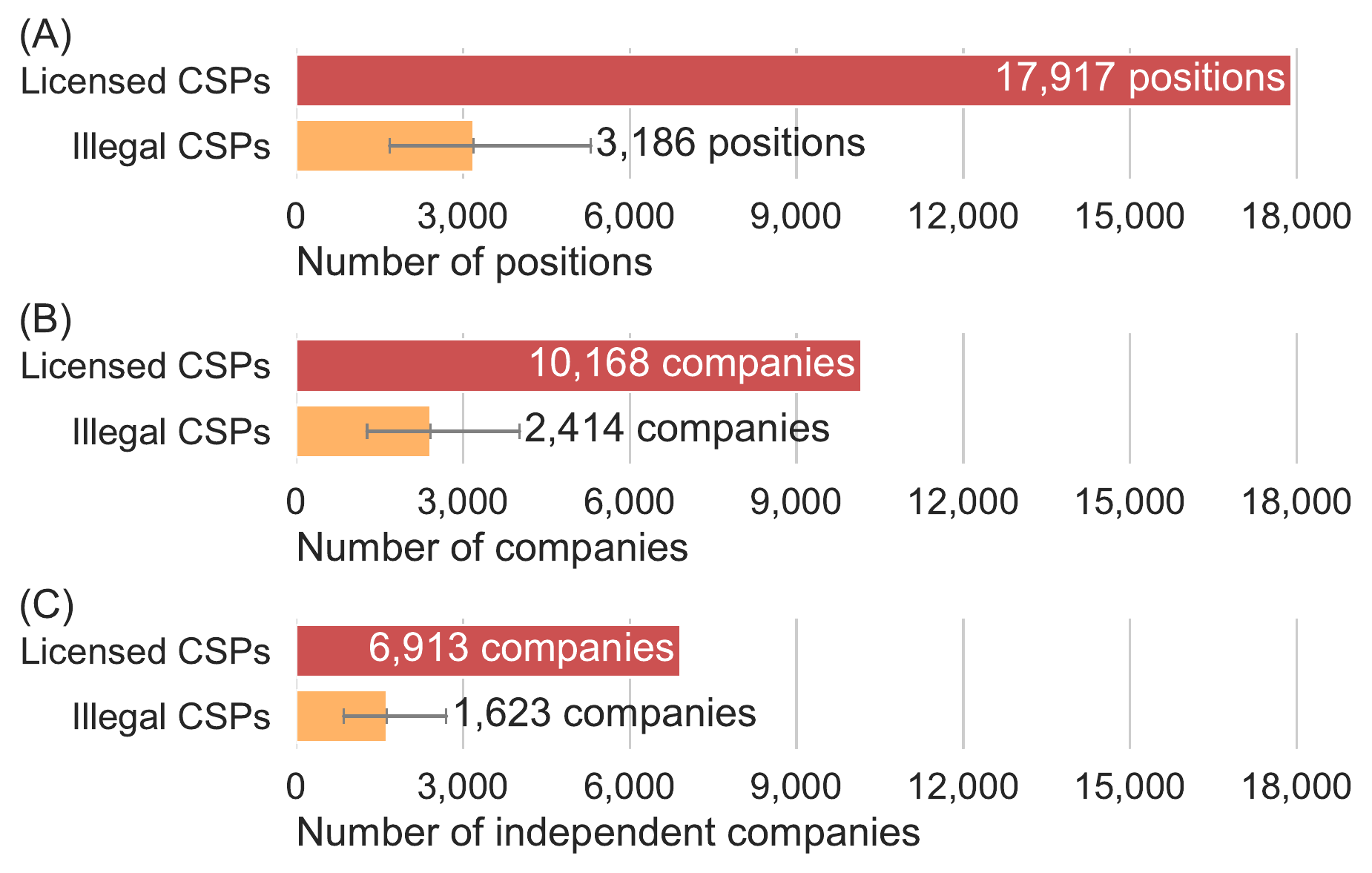}
    \caption{Number of (A) directorship position held (B) companies serviced (C) independent companies serviced by licensed CSPs (red) and illegal CSPs (orange).}
    \label{fig:summ_indep}
\end{figure}

Figure \ref{fig:summ_indep} visualizes the number of companies serviced by both licensed and illegal CSPs. Illegal CSPs operate on a much smaller scale. The average licensed CSP director manages 19.7 companies (and the median one 10 entities), but the average illegal CSP director manages only 7.9. In total, licensed CSP-directors hold 17,917 directorship positions in 10,168 companies. This implies that more than one CSP can hold a position in the same company. The 10,168 companies are owned by 6,913 unique owners (i.e., not all companies are independent). Illegal CSPs only hold 3,186 (1,680--5,295) directorship position in 2,414 (1,275-4,010) companies, of which 1,622 (856--2,685) are independent. This implies a 15\% (9--23\%) market share for illegal CSPs in terms of the number of positions, and a 19\% (11--28\%) market share for illegal CSPs in terms of (independent) companies serviced.

\subsection{Characteristics of the illegal CSP sector} 
\begin{figure}[h!]

    \includegraphics[width=0.7\textwidth]{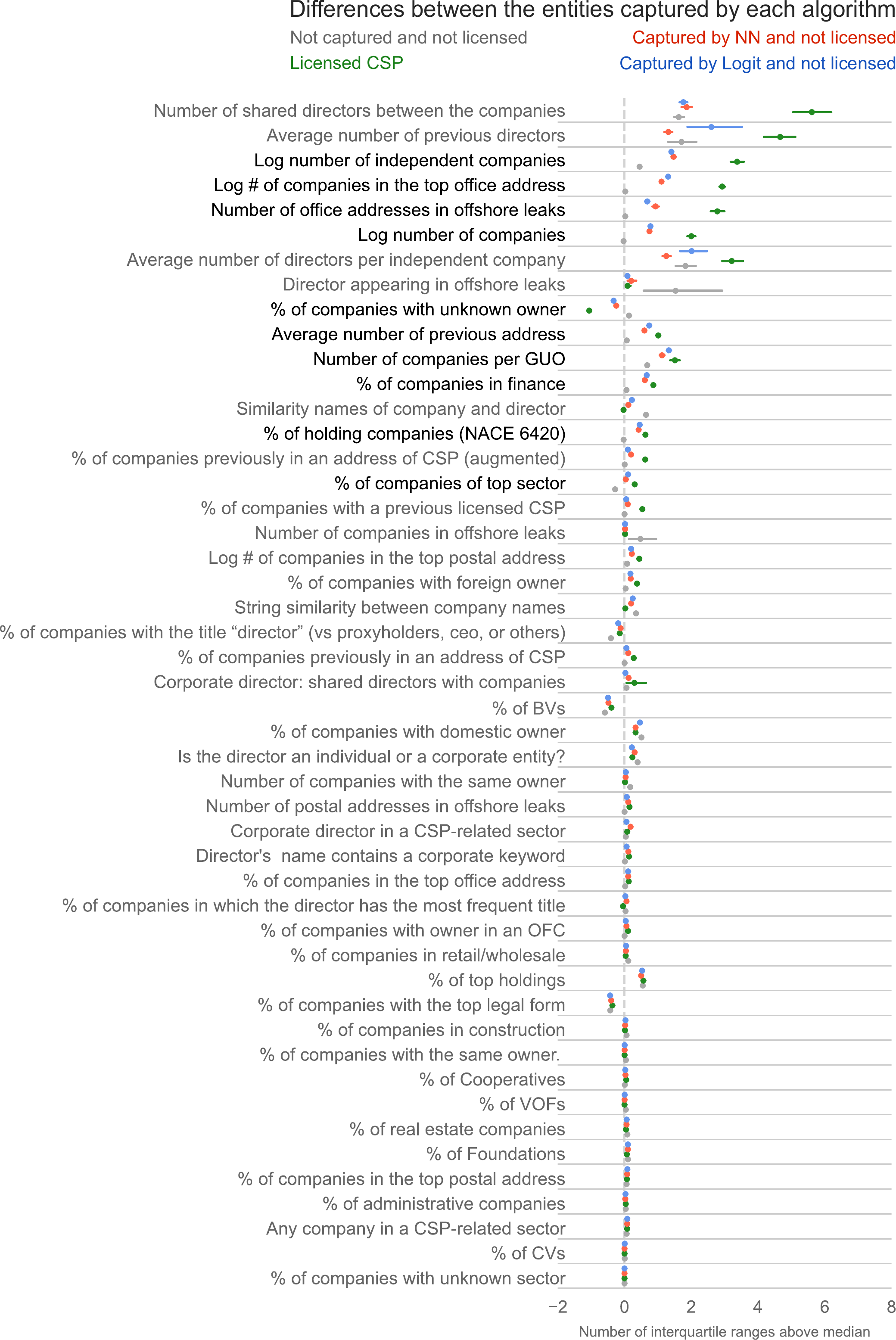}
    \caption{Mean normalized value for each variable, by group: licensed CSPs (green), non-licensed directors captured by the nearest neighbors algorithm (red), non-licensed directors captured by the logistic regression (blue), and non-licensed directors not captured by any algorithm (gray). Variables have been normalized to represent the number of interquartile ranges over the median value. Variables highlighted in black show different values for the flagged directors (red and blue) compared with non-flagged directors (gray). Confidence intervals are found via bootstrapping. See Sections \ref{sec:data} and \ref{sec:variables_detail} for a detailed explanation of the variables.}
    \label{fig:mean_variable}
\end{figure}

The classification models allow us to answer our main research question: ``what is the size of the illegal CSP sector in the Netherlands?''. To understand which features are correlated with potential illegal CSPs, we calculated, for each feature, the mean value for licensed CSPs, directors marked as CSPs by the nearest neighbors approach, and directors marked as CSPs by the logistic regression model. In order to allow for a more fair comparison between features, we normalized them by removing the median and scaling them using the interquartile range, an approach that is more robust to outliers. We show the average value of the normalized features in Figure \ref{fig:mean_variable}. We found that, compared with the directors not flagged by any algorithm and not licensed, some variables (labeled in black in Fig. \ref{fig:mean_variable}) seem to be different for the directors flagged by our algorithms. This was the case for the number of independent companies managed by the director, the number of companies in the most common office address, the number of addresses appearing in the offshore leaks, the number of total companies managed by the director, the share of companies with an unknown owner, the average number of previous addresses, the number of companies per independent company, and the share of companies in finance (especially if concentrated holding companies). Taken it together, the results paint a picture in which CSPs register several holding companies in one address. Interestingly, the owner is more likely to be known for companies managed by CSPs, which may reflect effective customer due diligence. It is worth noting that different models of illegality are possible (Section \ref{app:interviews}), and this heterogeneity may be masked by the average.

The use of a model-based classification algorithm (penalized logistic regression) allows us to directly understand which features are more important at predicting licensed CSPs (Fig. \ref{fig:logit}). We find that directors managing several financial companies (especially holding companies), with foreign owners (and especially those with owners in offshore financial center), and registered at the same address were more likely to be providing corporate services.
Conversely, directors managing partnerships (VOFs and CVs) belonging to the same group (high string similarity between company names, high number directors that are also owners) owned by several owners (either domestic or unknown) were more likely \textit{not} to be providing corporate services.
Interestingly, appearing in the offshore leaks had only a small effects, which may indicate that the supervision of the DNB is effective at preventing licensed CSPs to engage on illicit activities. 

\subsection{Validation and robustness tests}  \label{sec:results:small_entities}
We conducted two main validation and robustness tests. First, we ensured that we estimated the full extent of the illegal CSP-industry by employing two algorithms (nearest neighbors and penalized logistic regression), both tuned to reduce false negatives at the expense of increasing false positives. This validation test is detailed above, in sections \ref{sec:models} and Figure \ref{fig:flowchart}). 

Second, we investigated the effect of feature selection. To do so, we included a random subset of 80\% of the variables, and compared the overlap between the results found by our baseline algorithm (the one used throughout the paper) and the algorithm with 80\% of the variables included. We found a good overlap (75--80\%) between the directors flagged by our baseline and robustness algorithms (Section \ref{sec:robust_var_selection}).

\begin{figure}[h!]
    \includegraphics[width=0.5\textwidth]{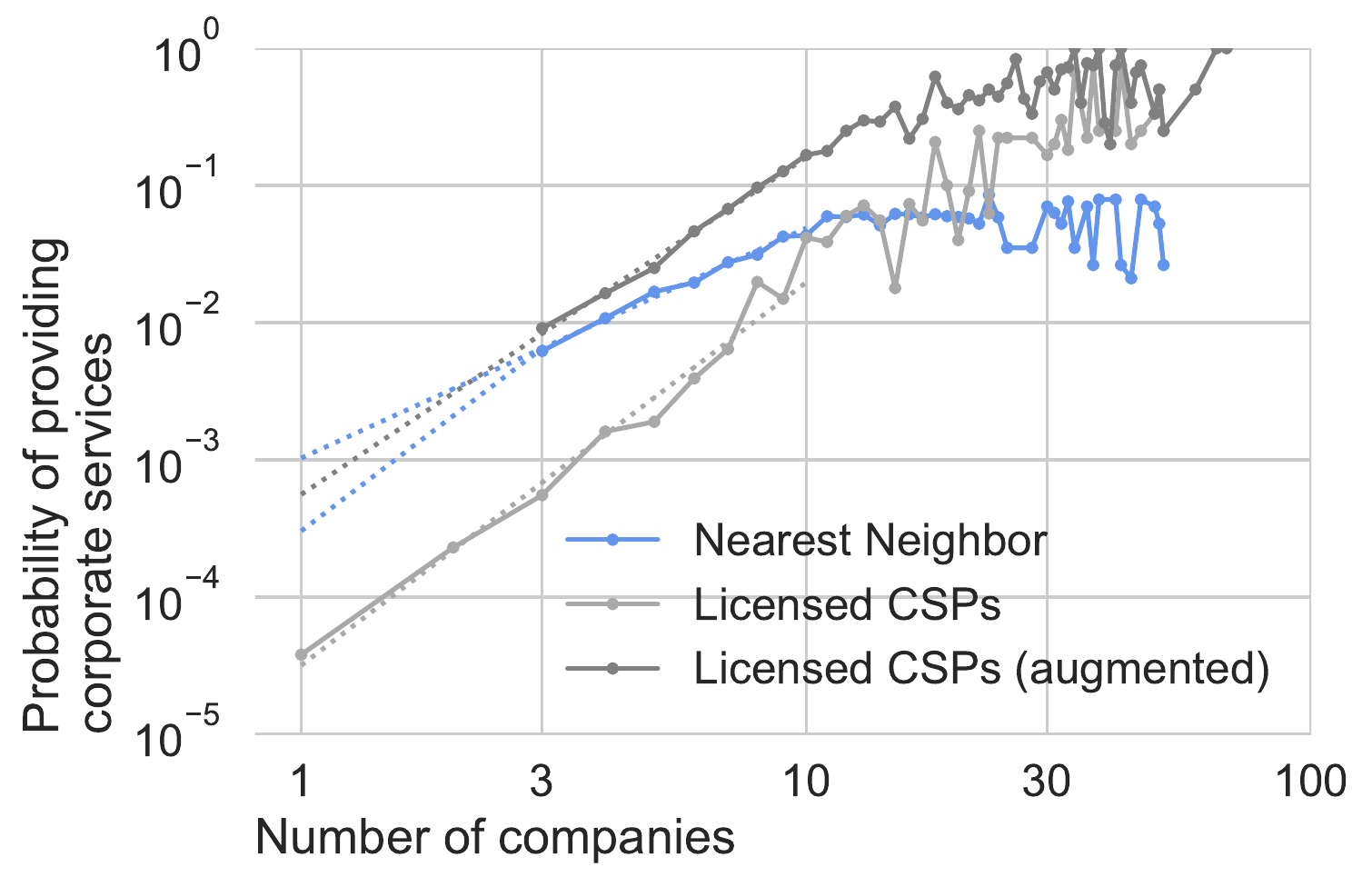}
    \caption{Extrapolation to smaller companies for the sample of licensed CSPs downloaded from the DNB (light gray), the full sample of licensed CSPs (dark gray), and the sample found with the nearest neighbor algorithm adjusted (blue). Note that the probability in the nearest neighbor sample has been adjusted using a uniform 11\% true positive rate, as found in the manually annotated sample. Dashed lines show a fitted linear model for directors managing 1--10 companies. For the nearest neighbor sample, a second extrapolation using the slope of the full sample of licensed CSPs is shown.}
    \label{fig:extrapolation}
\end{figure}

Third, we investigated the effect of the selection criteria explained in Section \ref{sec:data}, namely the exclusion of directors with one or two positions. To do so, we plotted the fraction of licensed CSPs as function of the number of independent companies managed by the directors (Fig. \ref{fig:extrapolation}). Around 50\% of the directors managing over 30 companies are licensed CSPs (some non-CSP directors also hold many positions, particularly those involved in fund management, real state, and in food processing). For directors managing less than 10 companies, the probability of providing CSPs decreases linearly (in log-log scale) as the number of managed companies decreases. Amongst directors managing three companies, only 1 in 100 directors is a licensed CSP (Fig. \ref{fig:extrapolation}). In the original list of licensed CSPs (i.e., not augmented using Orbis), we can see that the linear relationship continues for CSPs managing one or two companies. 
The linear decrease in the probability of acting as a CSP is also observed in the nearest neighbor sample (Fig. \ref{fig:extrapolation}, blue line). We can use this linear relationship to predict the potential number of directors managing one or two firms. Extrapolating using the decay in the sample identified by the nearest neighbor algorithm gives a higher bound of 1,645 directors (managing 2,046 companies). The true positive rate is likely to depend on the number of companies managed. We had scaled the entire distribution using a uniform 11\% true positive rate (the rate found via manual annotation). However, for directors managing over 30 companies, the true positive rate is likely to be higher (we see this behavior for the case of licensed CSPs in Figure \ref{fig:extrapolation}, gray lines). This implies that the decay should be more pronounced than the estimated using a uniform true positive rate. In order to find a lower bound, we extrapolated the sample identified by the nearest neighbor algorithm using the linear relationship found in the sample of licensed CSP. This gives an estimate of 820 directors (managing 1,112 companies). Including these directors in the sample, the market share (share of independent companies managed) of illegal CSPs would increase by 50--100\%, from 19\% to 23--27\%.

\label{sec:extrapolation}

\section{Conclusion} \label{sec:conclusion}
%Summary of the paper
Corporate services providers (CSPs) can facilitate economic crime by enabling the establishment of shell companies and obscuring the recipients or the nature of economic transactions. Such risks are higher for illegal (i.e. non-licensed) CSPs, as they circumvent the stricter regulation that applies to licensed CSPs. In this paper, we develop a method to estimate the size of the illegal CSP industry, finding illegal CSPs based on their similarity with licensed CSPs, where their similarity is computed based on their position within the network of directors, companies and addresses, and the characteristics of such entities.
The performance of our nearest neighbors method was validated in two ways. Firstly, we validate the precision by manually annotating a sample of the potential illegal CSPs. Using this method, we estimate that 11\% of the cases (330 cases extrapolated to the full sample) have a very high likelihood of providing illegal services. 
Secondly, we explore whether our method is capturing most illegal activity by fitting a penalized logistic regression to predict a new sample of potential illegal CSPs not previously flagged by the nearest neighbors algorithm. We again manually annotate a sample and estimate that 1.6\% (61 new cases) are highly likely to provide illegal services.

%Impact
Our analysis estimates that illegal CSPs represent 19\% of the market size---i.e., number of companies managed---and 31\% of all CSPs. These numbers increase to 23--27\%  and 38--51\% respectively when we extrapolate to include small illegal CSPs---those managing one or two companies. Our results have strong policy implications. A large share of the CSP industry is evading regulatory oversight and potentially facilitating other types of economic crime. Regulators could implement a red-flag system based on our network approach and actively detect illegal CSPs---especially with new data available on the ultimate beneficial owner of companies \cite{rijksoverheidUBOregisterFinancieleSector2019}.

%Future work
The analysis presented in this paper has a number of limitations that open up avenues for new research. The first limitation arises from the assessment of illegality based on the authors' knowledge instead of an audit process carried by a supervisory authority or fiscal crime unit. We are only able to identify entities with a high-risk of providing illegal services based on our manual annotation, and approximately 20\% of our observations were inconclusive. Future studies could partner with local authorities and implement our algorithm with two goals: (i) audit a sample of the potential candidates found in our analysis and establish true illegality. The audited CSPs could be fed back to the algorithms in order to improve their predictive power. (ii) use our approach to set up a red-flag system to continuously monitor and investigate illegal CSPs.  The design considerations for detecting illegal CSPs in a financial crime unit are likely different from the considerations of this paper. The objective of this paper has been to estimate the size of the illegal CSP industry, and as such we deliberately aimed at capturing most potential illegal CSPs at the expense of increasing the number of false positives. As a result, our true positive rate was relatively low, at around 11\%. In a situation where resources are limited, the true positive rate could be inflated for example by increasing in the nearest neighbors algorithm the minimum number of licensed CSP-neighbors necessary to flag a director as a potential illegal CSP.

Another limitation of our analysis that could be addressed in future research is that we flagged illegal CSPs based in their similarity with licensed CSPs. A cunning (unlicensed) CSP could mimic the behavior of directors not providing corporate services. For example, they could establish companies in many addresses, or use a network of front persons to become directors. Future investigations in collaboration with supervisory bodies could investigate if illegal CSPs mimicking ``normal'' directors can be detected. This could be achieved by leveraging data on previous fiscal investigations to train the algorithm, or by adding extra layers in the network connecting those companies---e.g., using transactions between the companies, or using the employment affiliation of the directors.

A third avenue for future research consists on applying our model to other domains. Our results show that the network structure holds predictive power over illegal activities. Similar approaches to identify economic crime could be used in a variety of fields. For example, it could be applied to improve the effectiveness of customer due diligence, a regulation by which financial institutions must verify the identity of their clients and detect potential risks. 
Financial institutions could use their knowledge on the network of previously risky clients to assess the risk of new customers.
In this case, the network would link clients with addresses, bank accounts and companies.
It could also be applied in suspicious transaction reporting (transactions between individuals potentially related to money laundering or terrorist financing). Financial institutions and criminal investigators could use the network of previous suspicious individuals to find similar individuals to audit. In this case, the network would link individuals through companies, addresses, family relations and financial transactions.

\section{ List of abbreviations}
\begin{itemize}
    \item CSP: Corporate Service Provider (providers of corporate service requiring a license, ``trustkantoren'' in Dutch).
    \item DNB: De Nederlandsche Bank (the Dutch central bank )
    \item NACE Rev. 2: Statistical classification of economic activities 
    \item Wtt 2018: Act on the Supervision of Trust Offices 2018 (``Wet toezicht trustkantoren'')
    \item Wwft: Dutch Act on the Prevention of Money Laundering and Financing on Terrorism (``Wet ter voorkoming van Witwassen en financieren van terrorisme'')
\end{itemize}

\section{Declarations}
\subsection{Availability of data and materials}
  The data that support the findings of this study are available from Orbis but restrictions apply to the availability of these data, which were used under license for the current study and are not publicly available. All other data-sets and replication code are included. 
  
\subsection{Competing interests}
  The authors declare that they have no competing interests.

\subsection{Funding}
    Part of the analysis was commissioned by the Dutch Minister of Finance and carried out by SEO Amsterdam Economics. The original analysis resulted in the report ``Illegale trustdienstverlening'', and was designed, conducted and interpreted with full independence from the funding body. All non-anonymized data collected during the original study was deleted.
  
\subsection{Author's contributions}
 JW and JGB developed the network analysis and interpreted the results. JGB run the analysis. MV conducted the interviews with representatives from the industry that guided the study. All authors wrote, read and approved the final manuscript.

\subsection{Acknowledgements}
    We appreciate the hard work of our research assistants Sarah Leuthold and Rachid Aguelmous downloading Orbis data and manually annotating the directors.
    We thank Melis van der Wulp, Cees Schaap and Johannes Hers for the useful insights and discussions, and two anonymous reviewers to help us clarify and strengthen the paper.

% if your bibliography is in bibtex format, use those commands:
\bibliographystyle{bmc-mathphys} % Style BST file (bmc-mathphys, vancouver, spbasic).
\bibliography{W_paper_trust.bib}      % Bibliography file

\clearpage
\appendix

\renewcommand{\thefigure}{A\arabic{figure}}
\setcounter{figure}{0}
\renewcommand{\thefigure}{A\arabic{table}}
\setcounter{table}{0}
\appendix

\section{Appendix}
\subsection{Supervision of CSPs by the DNB} \label{app:supervision}
The Dutch Central Bank (DNB) is responsible for the supervision of CSPs. The costs of supervision consist primarily of the payroll costs of the DNB staff concerned with overseeing compliance on the Act on the Supervision of Trust Offices 2018. Oversight consists of gathering data on CSPs, inspection visits and examining signals of illegal activities. The total costs of supervision are fully passed on the trust sector according to the principle ``the polluter pays''. As such, total costs are divided over all the legal CSPs i.e., the CSPs that hold a trust license. Costs per service provider are corrected for total earnings from corporate services, such that larger offices pay a larger share of total costs. 

\subsection{Data download} \label{app:orbis}
All data was collected in July 2020. We downloaded the following variables from Orbis. Not all variables were used in the analysis.

\begin{table}[h!]
\begin{tabular}{p{4.5cm}p{3.5cm}p{3cm}p{4.5cm}}
\toprule
\textbf{Addresses}                                            & \textbf{Legal events}                       & \textbf{Directors}                                & \textbf{Financials and ownership } \\                 \midrule 
Company name Latin   alphabet                        & BvD ID number                      & BvD ID number                            & BvD ID number                            \\
BvD ID number                                        & Legal events - Date                & DM-UCI (Unique Contact Identifier)       & P/L before tax-th USD Last avail. yr     \\
VAT/Tax number                                       & Legal events - Description         & DM-Full name                             & Total assets-th USD Last avail. yr       \\
Trade register number                                & Legal events - Source              & DM-Job title                             & Shareholders funds-th USD Last avail. yr \\
National legal form                                  & Legal events - Official identifier & DM-Current or previous                   & Number of employees-Last avail. yr       \\
Operating revenue   (Turnover)-th USD Last avail. yr & Legal events - Type                & DM-Corresponding BvDID (when applicable) & NACE Rev. 2, core code (4 digits)        \\
Address line 1-Local   Alphabet                      & Legal events - Details             &                                          & CSH - BvD ID number                      \\
Postcode-Local Alphabet                              &                                    &                                          & CSH - Name                               \\
City-Local Alphabet                                  &                                    &                                          & CSH - Country ISO code                   \\
Country ISO code                                     &                                    &                                          & CSH - Operating revenue (Turnover)-m USD \\
Street-Local Alphabet                                &                                    &                                          &                                          \\
Street number-Local   Alphabet                       &                                    &                                          &                                          \\
Address line 1-Local   Alphabet.1                    &                                    &                                          &                                          \\
Postcode-Local   Alphabet.1                          &                                    &                                          &                                          \\
City-Local Alphabet.1                                &                                    &                                          &                                          \\
Country ISO code-Local   Alphabet                    &                                    &                                          &                                          \\
Street-Local Alphabet.1                              &                                    &                                          &                                          \\
Street number-Local   Alphabet.1                     &                                    &                                          &                                          \\
PO Box-Local Alphabet                                &                                    &                                          &  \\ \bottomrule                                      
\end{tabular}
\caption{Variables downloaded from Orbis}

\end{table}

\subsection{Splitting addresses} \label{app:regex}
Orbis reports former addresses using the format ``Formely:Street Number|Postcode CITY'' (e.g., ``Formerly: Locatellikade 1|1076 AZ AMSTERDAM"). We removed ``Formerly: ;; and split the string using the horizontal bar. The part to the \textit{right} was used to obtain the postcode and city using the following regular expressions (run consecutively until one of them had a match): \Verb[fontshape=it]|/\d{4}\s?[A-Z]{2}/| (Dutch standard), \Verb[fontshape=it]|/([A-Z][A-HJ-Y]?\d[A-Z\d]? ?\d[A-Z]{2}|GIR ?0A{2})/| (UK standard), \Verb[fontshape=it]|/([A-Z]?-?\d{4,5})(-\d{4})*/| (US standard). The part to the \textit{left} was used to obtain the street name and streen number using the following regular expressions \Verb[fontshape=it]"^(.*?)\s+(\d*\w*(\-|\/)?\d+.*)$/", matching the standard address format in the Netherlands (e.g. NOTELAAR 12, with potential additions at the end), and \Verb[fontshape=it]"/^(\d*\w*(\-|\/)?\d+.*?)\s+(.*?)$/", matching primarily foreign address.



## A.4  Variables in the study

| Variable in code | Explanation in paper | Detailed explanation | Reason for inclusion | Expectation |
|---|---|---|---|---|
| **Characteristics of directors** | | | | |
| corp | Is the director an individual or a corporate entity? | 1 if corporate | Points to CSP | Positive |
| corp_services | Director's name contains a corporate keyword | 1 if director's name contains a corporate keyword: FINANCIAL,TAX, MANAGEMENT, TRUST, CORPORATE, CONSULTANT, ADVIES, ADVISORS , MANAGERS , SERVICES , FIDUCIA , ADMINISTRATION | Points to CSP | Positive |
| dir_nace_check | Corporate director in a CSP-related sector | The sector of the corporate director is in NACE codes: 8211, 6619, 6920, 7022, 6430 | Points to CSP | Positive |
| n_dir_icij | Director appears in offshore leaks | 1 if it appears in the offshore leaks | Illicit activities | Positive |
| **Characteristics of both directors and companies** | | | | |
| same_directors | Corporate director: shared directors with companies | For corporate directors: Number of shared directors between the companies and the director itself | Same corporate group | Negative |
| mean_sim_dir_comps | Name similarity between companies and director | Mean string similarity between the company names and the director's name | Same corporate group | Negative |
| dir_freq | % of companies with the title "director" (vs proxyholders, ceo, etc) | % of companies in which the directors holds the title "director" (vs proxyholders, ceo, or others) | Coherence in titles may point to CSP | None/Positive |
| top_title | % of companies in which the director has the most frequent title | % of companies in which the director has the most frequent title | Coherence in titles may point to CSP | None/Positive |
| number_shared_directors | Number of shared directors between the companies | Number of shared directors between the companies | Same corporate group | Negative |
| av_n_prev_dir | Average number of previous directors per company | Average number of previous directors per company | Real companies | Negative |
| frac_dir_prev_rto4 | % of companies with a previous licensed CSP | Share of companies with a previous director connected to a licensed CSP | Previously managed by CSP, not anymore | Positive |
| **Characteristics of the addresses** | | | | |
| log_n_comps_top_office | Log # of companies in the top office address | Log10 total number of companies registered in the most common office address. All companies are considered. | Points to licensed CSP | Positive |
| log_n_comps_top_postal | Log # of companies in the top office address | Log10 total number of companies using the most common postal address | Points to licensed CSP | Positive |
| office_top | % of companies in the top office address | Fraction of independent companies (amongst the ones managed by the director) registered in the most common office address | Points to CSP | Positive |
| postal_top | % of companies in the top postal address | Fraction of independent companies (amongst the ones managed by the director) using the most common postal address | Points to CSP | Positive |
| frac_add_prev_trust | % of companies previously in an address of CSP (augmented) | Fraction of independent companies (amongst the ones managed by the director) registered in an address present in the augmented (DNB+Orbis) list of licensed CSPs | Previously managed by CSP, not anymore | Positive |
| frac_add_prev_trust_off | % of companies previously in an address of CSP | Fraction of independent companies (amongst the ones managed by the director) registered in an address present in the DNB list of licensed CSPs | Previously managed by CSP, not anymore | Positive |
| frac_add_off_icij | Number of office addresses in offshore leaks | Number of office addresses present in the offshore leaks | Illicit activities | Positive |
| frac_add_post_icij | Number of postal addresses in offshore leaks | Number of postal addresses present in the offshore leaks | Illicit activities | Positive |
| av_n_prev_add | Average number of previous address | Average number of previous address | Real companies | Negative |
| **Characteristics of both owners and companies** | | | | |
| log_total_guo | Log number of independent companies | Log10 number of independent companies (i.e., with different global ultimate owners) | Points to CSP | Positive |
| av_directors_comps | Number of directors per independent company | Average number of directors per independent company (i.e. global ultimate owner) | Independent companies | Negative |
| av_companies_guo | Number of companies per independent company | Average number of companies per independent company | Same corporate group | Negative |
| **Characteristics of the companies** | | | | |
| dir_in_comp_nace_check | Any company in a CSP-related sector? | The director is an individual serving in a board of a company with sectors 8211,6619,6920,7022,6430 | Points to CSP | Positive |
| log_n_comp | Log number of companies | Log10 number of companies | Points to CSP | Positive |
| mean_sim | String similarity between company names | String similarity between company names | Same corporate group | Negative |
| nace_finance | % of companies in finance | Share of companies in sectors NACE 64, 65, 66 | Financial activities | Positive |
| nace_holding | % of holding companies (NACE 6420) | Share of companies in sector NACE 6420 | Financial activities | Positive |
| nace_real_estate | % of real estate companies | Share of companies in sector NACE 68 | Financial activities | Positive |
| nace_top_holding | % of top holdings | Share of companies in sector NACE 70 | Financial activities | Positive |
| nace_adminis | % of administrative companies | Share of companies in sectors NACE 69, 82 | Financial activities | Positive |
| nace_unknown | % of companies with unknown sector | Share of companies with unknown sector | Financial activities | Positive |
| nace_retail_wholesale | % of companies in retail/wholesale | Share of companies in sectors NACE 45, 46, 47 | Productive activities | Negative |
| nace_construction | % of companies in construction | Share of companies in sectors NACE 41, 42, 43 | Productive activities | Negative |
| top_nace | % of companies of top sector | Share of companies in the most common sector | Coherence in activities may point to CSP | None/Positive |
| legal_bv | % of BVs | Share of companies in the legal type: BV | Limited liability | None/Positive |
| legal_found | % of Foundations | Share of companies in the legal type: Foundation | Allow obscuring of ownership | None/Positive |
| legal_vof | % of VOFs | Share of companies in the legal type: VOF | Partnership | Negative |
| legal_coop | % of Cooperatives | Share of companies in the legal type: Cooperative | Partnership | None/Negative |
| legal_cv | % of CVs | Share of companies in the legal type: CV | Partnership | None/Negative |
| top_legal | % of companies with the top legal form | Share of companies belonging to the most common legal type | Coherence in types may point to CSP | None/Positive |
| frac_comp_icij | Number of companies in offshore leaks | Number of companies in offshore leaks | Illicit activities | Positive |
| **Characteristics of the owners** | | | | |
| same_guo | Number of companies with the same owner | Number of companies with the same owner. | Same group | Negative |
| same_guo_frac | % of companies with the same owner. | % of companies with the same owner. | Same group | Negative |
| no_guo | % of companies with unknown owner | % of companies with unknown owner | Obscuring ownership | Positive |
| nl_guo | % of companies with domestic owner | % of companies with domestic owner | Easier to track | Negative |
| foreign_guo | % of companies with foreign owner | % of companies with foreign owner | Hard to track | Positive |
| ofc_guo | % of companies with owner in an OFC | % of companies with owner in an OFC | Harder to track | Positive |

## A.5 Codebook for manual annotation

**Manual annotation of directors**

*Please treat the data associated as confidential.*

We have developed a method to identify directors that may be providing illegal trust services. A director is providing trust services when she/he is the director of several companies on behalf of a third party. This implies that the director is not the owner of all the companies. Moreover, being the director of several companies is totally legitimate as long as you have expertise in the field of operations of the company—e.g. you can be a director of Coca-Cola, Philips and Nestle, since those are large companies requiring of the same type of expertise. Trust services are thus provided when the director is only a director "on paper", managing the accounts of the company but not doing much more.

A director can be a firm, it does not need to be an individual. In this case the firm must be independent from the companies where it acts as a director. For example, it is okay if "Nestle Dir" is the director of 10 subsidiaries of "Nestle". In the case of firms, a firm is also providing illegal trust services if it provides an address to another company and some other additional services (e.g. handling the tax returns, the accounts, etc).

Finally, a director is providing **illegal** trust services if it not registered with the DNB, which can be checked on their website. This implies that if Mr. X is working for INTERTRUST (a registered trust service provider), Mr. X is most likely providing legal trust services.

The goal of this step of the project is to understand the accuracy of our method. For this, we are providing a spreadsheet with 100 directors and the companies where (s)he is the director. The goal is to manually annotate as many as possible, trying to find if the director is providing trust services (by googling, using linkedin, google maps, Drimble, Orbis, etc), and if those trust services are legal or illegal (by checking the DNB website).

**Code book (examples in the following pages):**

| Question | Practical tips | Coding book |
|---|---|---|
| Q1: Can you find the director online? (who is the main employer of the director)? | Remove middle names, add "LinkedIn" to the query. | 0 (no), 3 (yes)<br><br>**Copy URLs in the spreadsheet** |
| Q2: Is the director employed by a registered trust service provider (if individual) or itself a registered trust service provider (if corporate director). | If you find the employer of the director, check here if it is registered | -9 (not applicable), 0 (no), 3 (yes)<br><br>**If yes don't answer the questions below** |
| Q3: Is the company employing the directors a:<br>Q3a: Private equity or investment firm<br>Q3b: Real estate firm<br>Q3c: Corporate service provider (tax/domiciliation/accountancy/advice) | | -9 (not applicable), -1 (Unknown), 0 (no), 3 (yes) |
| Q4: For the companies where the director has a position: Does the director seem to have expertise in that area? | e.g. A director works for several companies involved in IT. | -9 (not applicable), -1 (unknown), 0 (no), 1 (probably no), 2 (probably yes), 3 (yes) |
| Q5: For the companies where the director has a position: Are (most of) the companies' part of the same corporate group? | The variable Owner ID gives you some information, but not always correct, and you can look up the companies in https://drimble.nl | -9 (not applicable), 0 (no), 1 (probably no), 2 (probably yes), 3 (yes) |
| Q6: For the companies where the director has a position: Do they have a website? | | -9 (not applicable), 0 (no), 1 (mostly no), 2 (mostly yes), 3 (yes)<br><br>**Copy URLs in the spreadsheet** |
| Q7: (only if previous questions are mostly unknown): For the companies where the director has a position: Who were the previous directors of the companies? | e.g. https://drimble.nl/bedrijf/laren/17461871/maida-investment-bv.html | -9 (not applicable), -1 (unknown)<br>RTO (person working for an RTO, or an RTO itself), Other corporate (Other corporate–e.g. Maida), Other |
| Q8: Any other relevant information? | Anything that you find when investigating the directors and did not fit in the categories. | |

**Example 1:**
Ms. DIRECTOR is the director of 11 companies, here are the top 8:

| Type | Company/Director Name | Company ID | Owner ID | Office address | N. companies in office address | Postal address | N. companies in postal address | Postbus | NACE sector | Legal form | Country of owner | Number of directors in company |
|---|---|---|---|---|---|---|---|---|---|---|---|---|
| DIRECTOR | | NaN | NaN | | NaN | NaN | NaN | NaN | NaN | NaN | NaN | NaN |
| Comp1 | | | KONING | | 100.0 | 0 | NaN | NaN | 6420.0 | Private limited liability company - BV | NL | 1.0 |
| Comp2 | | | KONING | | 100.0 | 0 | NaN | NaN | 6420.0 | Private limited liability company - BV | NaN | 1.0 |
| Comp3 | | | KONING | | 100.0 | 0 | NaN | NaN | 6519.0 | Foundation | NaN | 3.0 |
| Comp4 | | | KONING | | 100.0 | 0 | NaN | NaN | 6831.0 | Private limited liability company - BV | NL | 1.0 |
| Comp5 | | | KONING | | 100.0 | 0 | NaN | NaN | 6920.0 | Private limited liability company - BV | NL | 1.0 |
| Comp6 | | | KONING | | 100.0 | 0 | NaN | NaN | 6420.0 | Private limited liability company - BV | NaN | 1.0 |
| Comp7 | | | KONING | | 100.0 | 0 | NaN | NaN | 7022.0 | Private limited liability company - BV | NL | 3.0 |
| Comp8 | | | KONING | | 100.0 | 0 | NaN | NaN | 7010.0 | Private limited liability company - BV | NaN | 1.0 |

You can see in the data that the owner of those companies is often missing, that they are all registered in the same office address and that there are 100 companies registered in the same address. The sector of many companies is 64/66/69/70, which are codes related to financial and support activities. Most of the companies where she is a director are BVs and generally she is the only director of the company.

Q1: If we search her on linkedin (using the query "▚▚▚▚▚▚▚▚▚▚▚▚▚▚▚ linkedin" on google is usually the fastest way), we can find her page, which shows that she was the owner of a trust management (▚▚▚▚▚▚▚▚▚▚) and the owner of ▚▚▚.

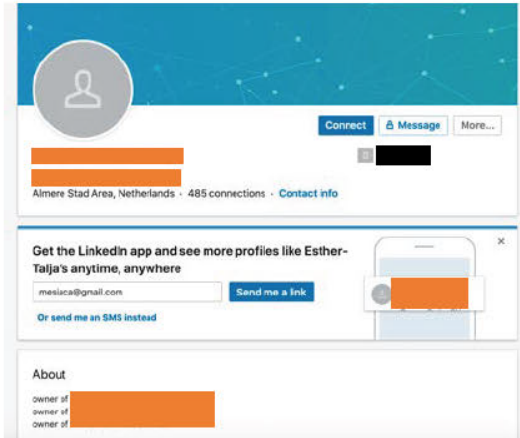

Q2: Checking the DNB register we find that ▚▚▚ is registered. We stop here



\subsection{Knowledge graph} \label{sec:neo4j}
\begin{figure}[h!]

    \includegraphics[width=\textwidth]{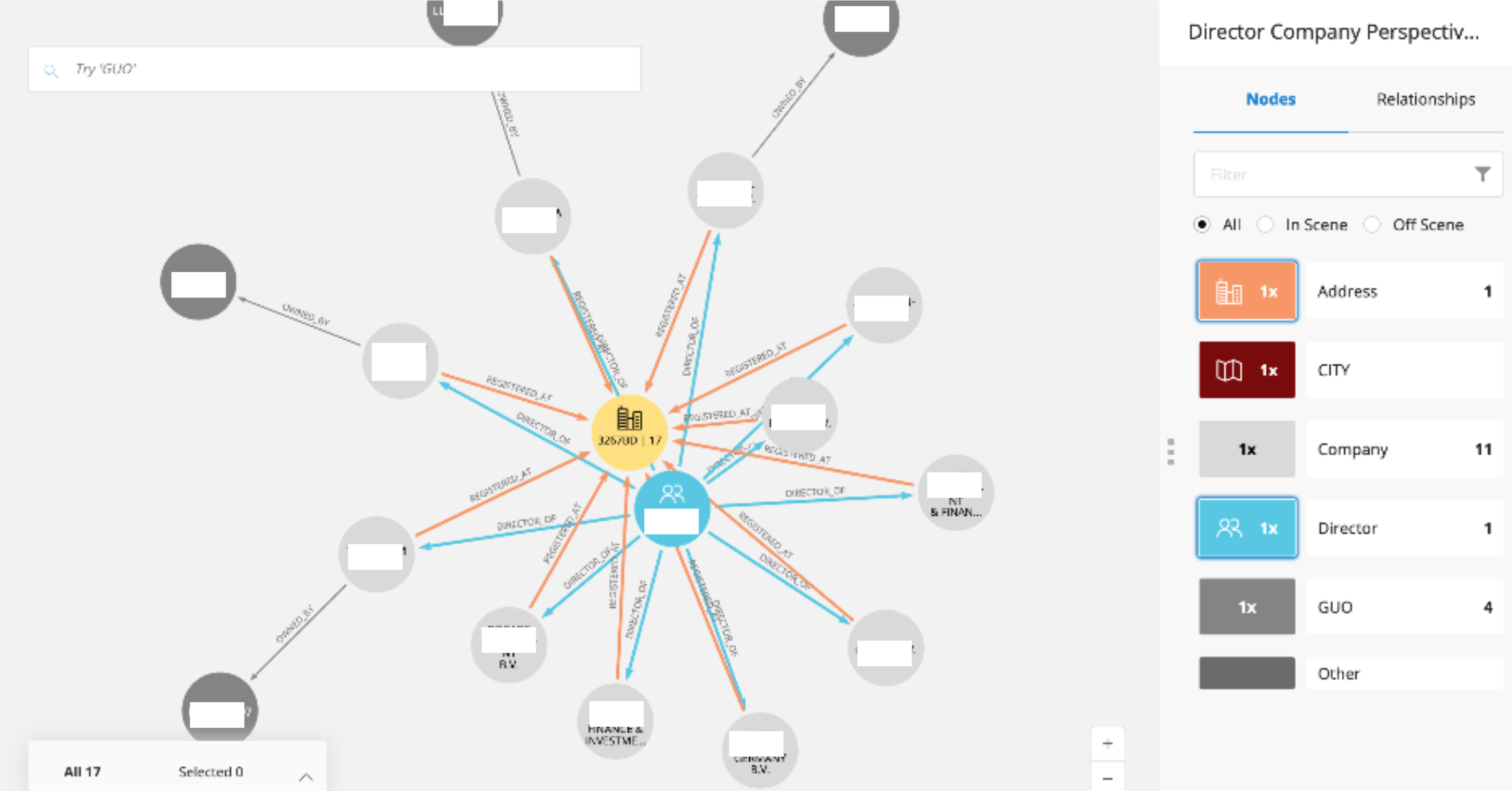}
    \caption{Example of a potential illegal CSP, visualized in Neo4J Bloom. Note that all companies (gray circles) are managed by the same director (blue circle), registered in the same address (yellow circle), and owned by different owners (dark gray circles). The director was afterward found in online sources. }
    \label{fig:neo4j}
\end{figure}

We developed a visualization of the network using the graph database Neo4J, and the visualization Neo4J Bloom. A potential illegal CSP is included in Fig. \ref{fig:neo4j}.

\subsection{Robustness tests: Variable selection}\label{sec:robust} \label{sec:robust_var_selection}
\begin{figure}[h!]

    \includegraphics[width=0.5\textwidth]{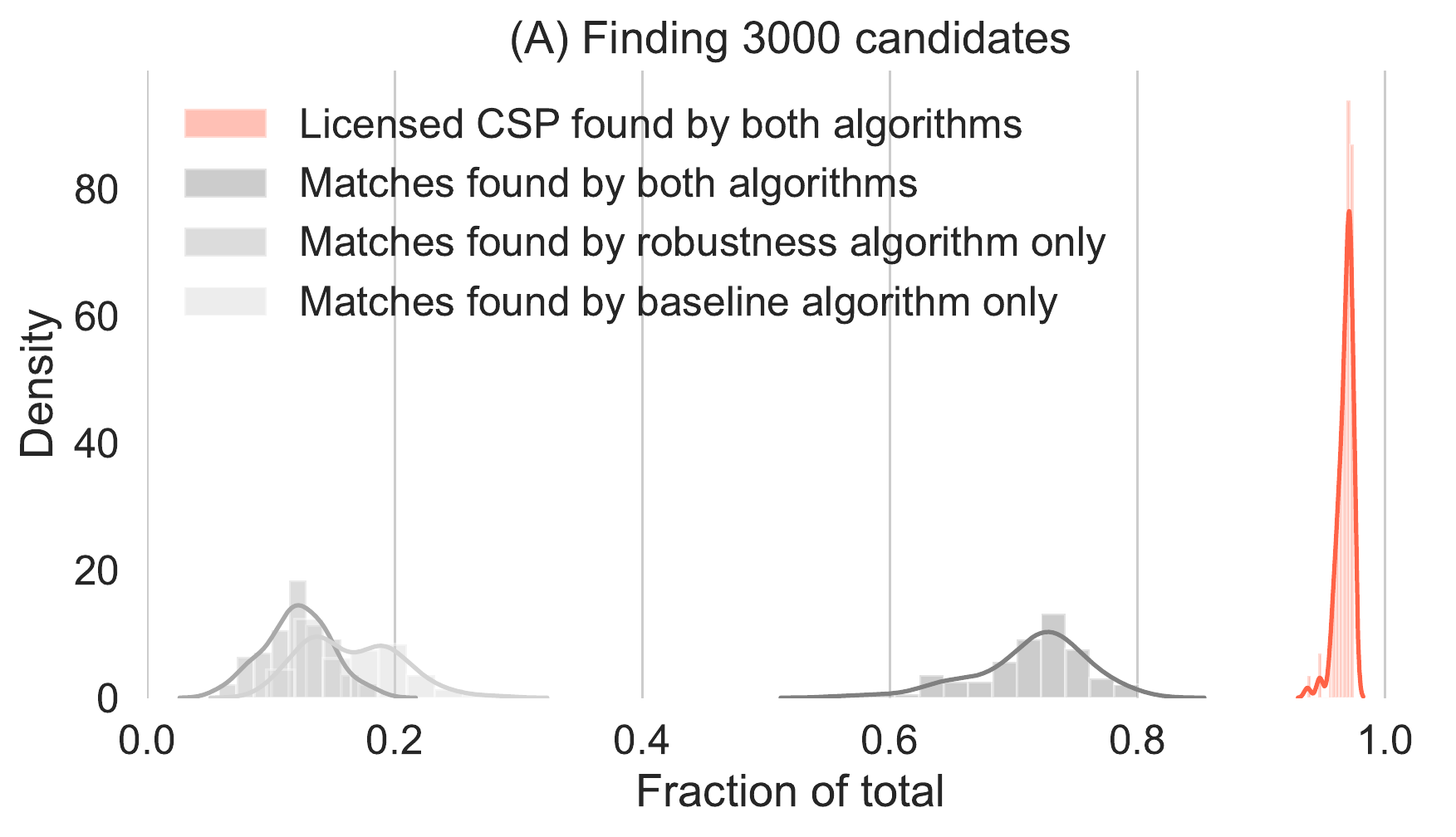}
    \includegraphics[width=0.5\textwidth]{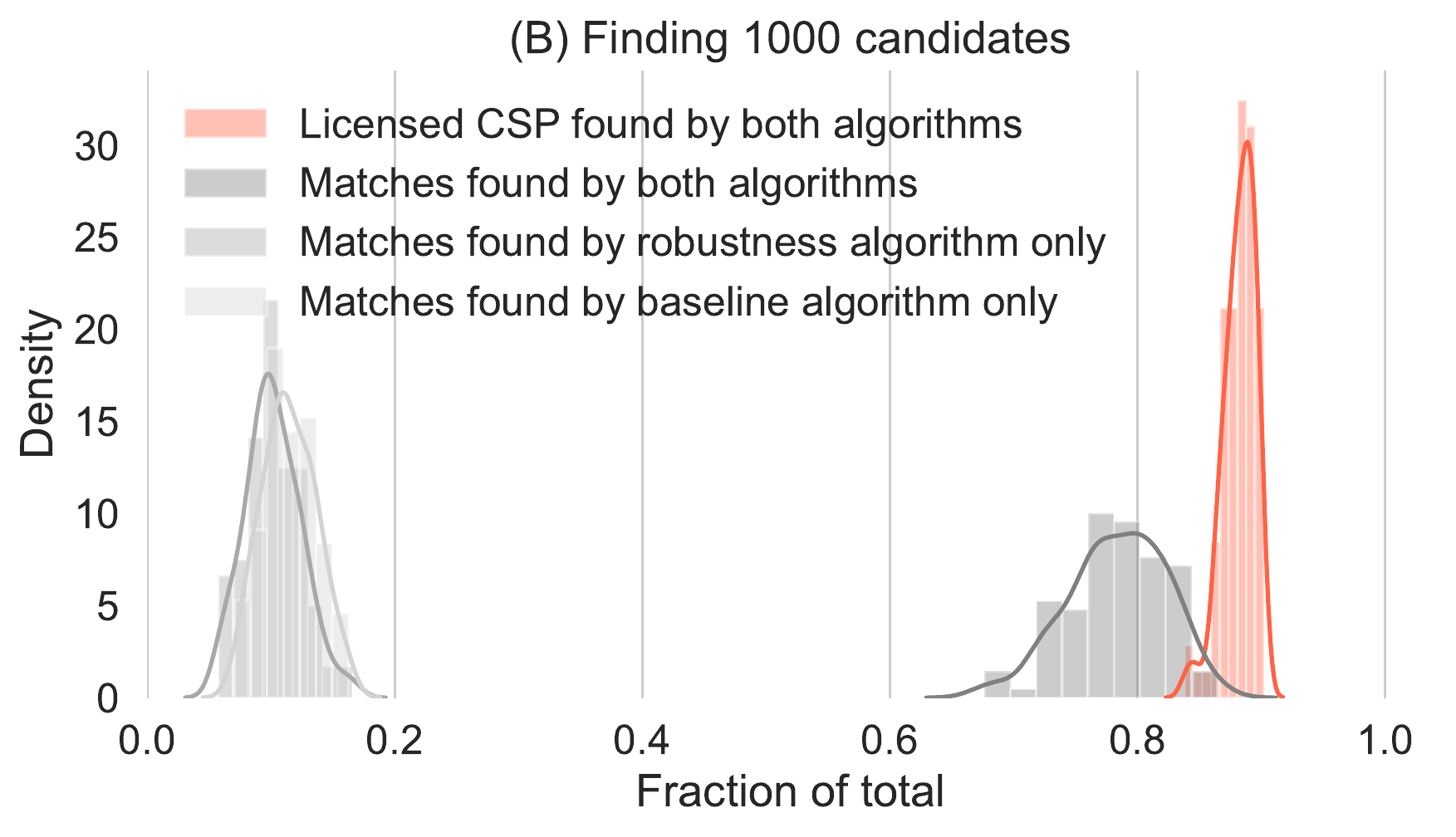}
    \caption{Effects of feature selection. The nearest neighbor algorithm selects directors close to at least (A) three (B) nine licensed CSPs. Note that both algorithms are able to find the majority of licensed CSPs (the red distribution peaks at around 90-95\%), and both algorithms find a similar set of directors (darker gray).
    }
    \label{fig:var_selection}
\end{figure}

We test the effect of feature selection in the nearest neighbors algorithm. To do so, we include a random subset of 80\% of the variables, and compared the overlap between the results found by our baseline algorithm (the one used throughout the paper) and the algorithm with 80\% of the variables included. We run this procedure 100 times and visualize the results in Figure \ref{fig:var_selection}.

We find a good overlap between the baseline and the robustness algorithm (Fig. \ref{fig:var_selection}). When the algorithm is set to keep directors near three or more licensed CSPs---our choice in the paper, returning approximately 3,000 candidates on top of the 909 licensed CSPs---both algorithms find over 95\% of the licensed CSPs, and both algorithms agree on 75\% of the directors marked as CSPs. When the algorithm is set to keep directors near nine or more licensed CSPs---resulting in approximately 1,000 candidates---both algorithms find 90\% of licensed CSPs, and both algorithms agree on 80\% of the directors marked as CSPs.

\subsection{Interviews} \label{app:interviews}
In order to construct the indicators, information was gathered on the characteristics of different kinds of illegal CSPs and how they differ from legal CSPs. The information was gathered by holding semi-structured interviews with legal and illegal CSPs. In the interviews, CSPs described their business model and more broadly the way they operate. From these interviews, the research team distilled indicators that are related to certain types of illegal CSPs. 

There are different ``models'' of illegal CSPs, that are also associated with different characteristics. For example, in order to keep out of sight of DNB and other authorities, illegal CSPs can choose to limit the scale on which they operate, i.e. limit the number of clients served. Alternatively, illegal CSPs can choose to limit the range of corporate services they offer, such that they cannot be directly linked to the full range of corporate services. For the corporate services they do not offer themselves, they can put forward a frontman or form a network with other service providers, such that the client still has access to the full range of corporate services. By dividing illegal corporate services over different players, illegal CSPs can serve a larger number of clients while still limiting the risk of being detected. 

Illegal CSPs face a trade-off between the scale of operation on the one hand, and the range of corporate services they personally offer on the other. Both scale and range of services typically result in higher profit, as well as a higher risk of being detected. According to microeconomic theory, illegal CSPs seek an equilibrium where scale and range are adjusted such that the utility of the illegal CSP is maximized.

As the preferences of illegal CSPs differ, so does their equilibrium outcome in the utility function. As such, we find different models of illegal CSPs that differ in scale and the range of services provided. For example, the CSP can limit the range of services personally provided in order to limit the detectability of the illegal corporate services. The CSP can form different networks with different facilitators in order to limit detectability. They can for example work together with different domiciliary service providers, such that domiciliation is fragmented over different addresses. As such, the directors themselves are the only recurring factor in the networks that they form with the other facilitators. The director is registered as a director of several entities and can therefore also be identified. 
The network becomes more difficult to detect if the CSPs uses frontmen. The CSPs cannot be directly affiliated with the network, because their own name does not appear in the registration of the entities. In practice, the role of the service provider remains the same: to act as a director through frontmen. The purpose of this model is again to conceal the links to the actual CSP. CSPs can recruit directors from various networks. The less the different networks are connected to each other, the more difficult it is to detect connections between the different networks. 
A third option is the ``self-coordinating straw man'' model. In this model, directors form a network in which they act as front men for each other. In one of the interviews, one respondent refers to a case in which former employees of trust offices form such a ‘network in a network’ i.e. act as a director in each other’s networks on a rotating basis.

\subsection{False negatives licensed CSPs}

\begin{figure}[h!]
    \includegraphics[width=0.5\textwidth]{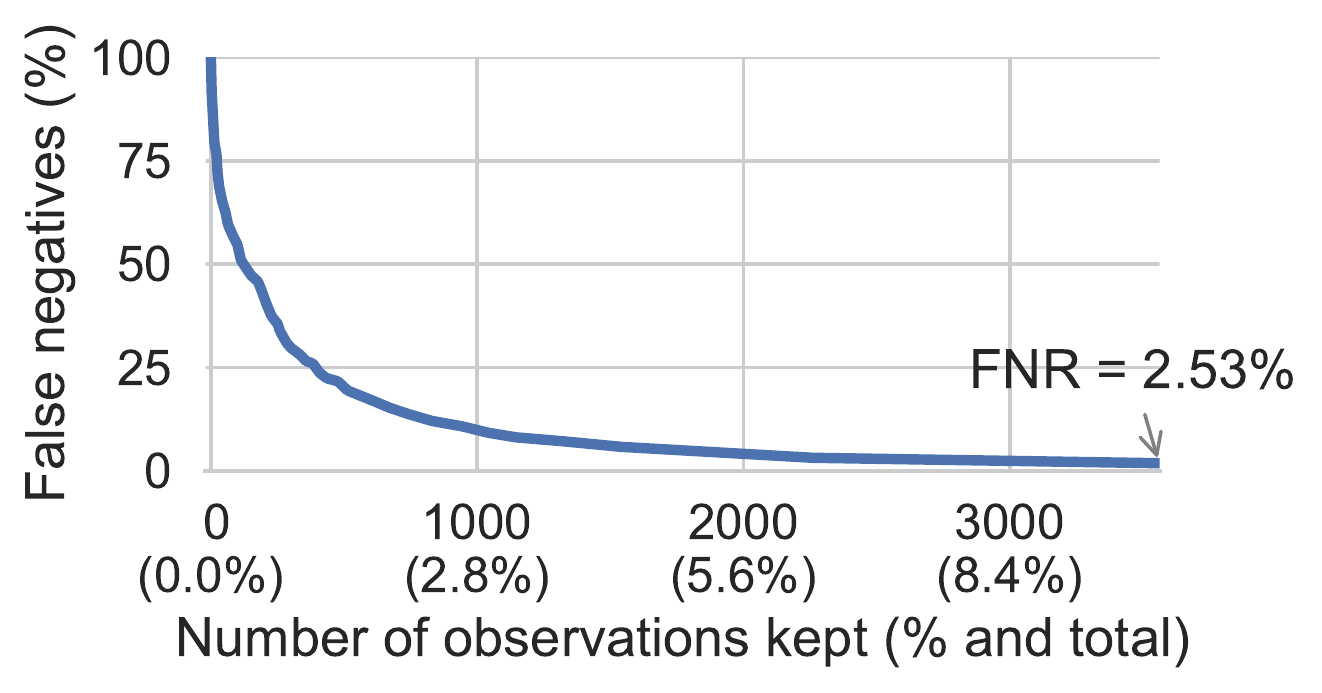}
    \caption{False negative rate (licensed CSPs not found by the nearest neighbors algorithm) as a function of the number of observations kept by the algorithm. The false negative rate (FNR) for our chosen sample (where observations near at least three licensed CSPs are kept) is annotated.}
    \label{fig:app_roc}
\end{figure}

\clearpage
\subsection{Effect sizes in the logistic regression}

\begin{figure}[h!]

    \includegraphics[width=0.8\textwidth]{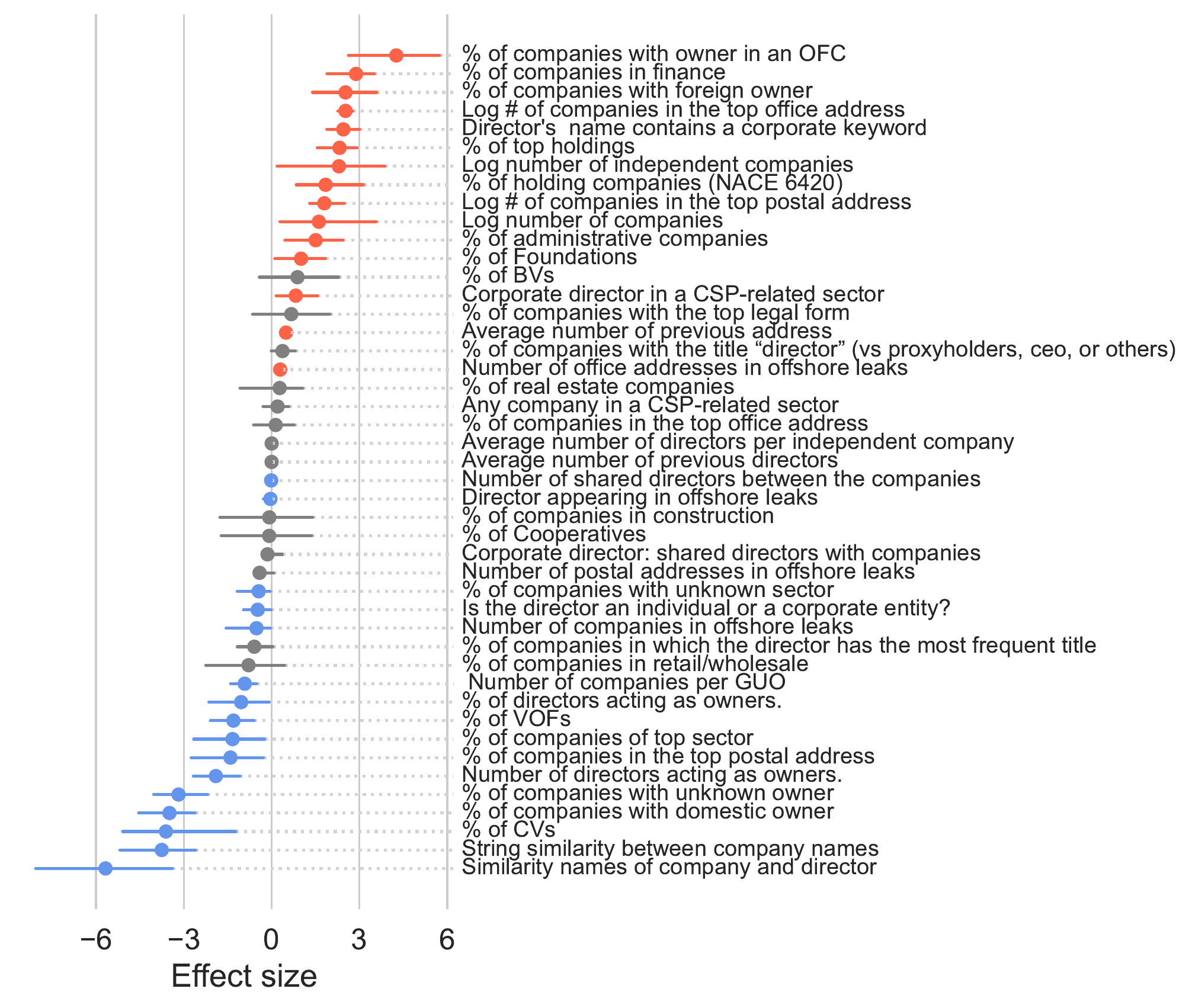}
    \caption{Effect size of each variable (feature) in the logistic regression. Confidence intervals are found via bootstrapping. Features with significant positive (negative) effect sizes are colored in red (blue). See Sections \ref{sec:data} and \ref{sec:variables_detail} for a detailed explanation of the variables.}
    \label{fig:logit}
\end{figure}

% \end{backmatter}
\end{document}